# International migration and dietary diversity of left-behind households: evidence from India

Pooja Batra

Jindal Global Business School, Sonipat, India

Email: f15poojab@iimidr.ac.in

Ajay Sharma

Indian Institute of Management Indore, Indore (India)

Email: ajays@iimidr.ac.in

Abstract

In this paper, we analyse the impact of international migration on the food consumption and dietary diversity of left-behind households. Using the Kerala migration survey 2011, we study whether households with emigrants (on account of international migration) have higher consumption expenditure and improved dietary diversity than their non-migrating counterparts. We use ordinary least square and instrumental variable approach to answer this question. The key findings are that: a) emigrant households have higher overall consumption expenditure as well as higher expenditure on food; b) we find that international migration leads to increase in the dietary diversity of left behind households. Further, we explore the effect on food sub-group expenditure for both rural and urban households. We find that emigrant households spend more on protein (milk, pulses and egg, fish and meat), at the same time there is higher spending on non-healthy food habits (processed and ready to eat food items) among them.

**Correspondence address:**

**Ajay Sharma**, J-206, Academic Block, Indian Institute of Management Indore, Prabandh Shikhar, Rau-Pithampur Road, Indore, M.P. (India) - 453556. Ph: +91-7312439622. E-mail: ajays@iimidr.ac.in; ajaysharma87@gmail.com.

**Conflict of Interest:** The authors do not have any conflict of interests to report.

**Data Availability Statement**

The data that support the findings of this study are openly available in MIGRATION SURVEY DATA, Centre for Development Studies, Thiruvananthapuram, Kerala, India at [http://cds.edu/research/ru/migrationresearch/migration-survey-data/, Last accessed on Oct 31, 2020]

1. Introduction

Sustainable Development Goals (SDGs) 2 and 3 are targeted towards ending hunger and ensure healthy lives and promote well-being for all ages. Since the goal setting in 2015, the progress towards achieving these goals has been really slow as the per the recent *The State of Food Security and Nutrition in the World 2020 report* (World Health Organization, 2020). The estimates based on this report suggests that around 3.1 billion people around the world do not have access to affordable and healthy diet.

In the literature on nutrition and food security, there are several pathways suggested to improve nutritional and food security for households ranging from agricultural production to women empowerment to farm and non-farm employment (Kadiyala et al., 2014). Out of these channels, non-farm employment remains and important channel in improving the well-being of households. Sadiddin et al. (2019) document how the incidence of food insecurity can be one of the key factors in inducing migration and non-farm employment across Sub-Saharan Africa. For a large population in developing countries, engaging in non-farm employment through mobility – commuting and migration remains an important channel in improving the well-being of left-behind households (Sharma and Chandrasekhar, 2016; Azzarri and Zezza, 2011; Kim et al., 2019). There were an estimated 244 million international migrants in 2015 who moved out to diversify their livelihood (Skeldon, 2018). Further, in countries with very less regional heterogeneity (in terms of economic growth and development), international migration remains the only viable option for diversification of income and upward mobility. This fact can be supported by the high share of international migration from small countries (in West Indies island countries, international migration rate varies from 66 percent to 93 percent (Guyana (93 percent), Haiti (75 percent), Trinidad and Tobago (68 percent each) and Barbados (66 percent)) (World Bank, 2016). Further, for the smaller countries, international remittances account for a significant share of their gross domestic product (Tajikistan (42 percent), the Kyrgyz Republic (30 percent), Nepal (29 percent), Tonga (28 percent), and Moldova (26 percent)).

Given this background, this study contributes to the literature on migration and its impact on increased well-being and food security of the left-behind household. For our study, we consider the state of Kerala in India that remains one such region where incidence of international migration is one of the highest in the country. This setting provides us an interesting window to analyse the returns to international migration in terms of impact on well-being indicators such as consumption expenditure, food, and non-food expenditure as well as nutritional intake and food security indicators like dietary diversity.

With a focus on understanding the impact of international migration on consumption expenditure and dietary diversity of the left-behind households through various channels, we draw from different strands of literature.

First strand of literature argues that irrespective of the destination (domestic or abroad), migration is viewed as an "indirect mean to sustaining livelihood", usually undertaken by households or individuals to diversify income, reduce economic shocks, minimize risks and maximize economic welfare (Stark and Levhari, 1982; Stark and Bloom, 1985). As per the New Economics of Labor Migration (NELM), migration is a choice (decision) which the households make as a livelihood strategy to diversify their source income as a means to "maintain, secure and improve their livelihoods" which enhances their standard of living (de Haas, 2010). Thus, both internal and international migration aim to achieve the same end, that is, of broadening horizons of livelihoods.

Second strand of literature argues that, for the migrant sending households, the main impact of migration is threefold (Zezza et al., 2011). First, there is possibility to receive remittances from the migrant that can have direct and indirect effects on consumption related expenditure (Isoto and Kraybill, 2017). Second is the informational impact of migration. Due to migration, the household is exposed to both positive and negative aspects of the outside world. This can have positive (improvement in educational outcomes and aspirations) or negative (bad food habits etc.) impact on the members of households (Kandel and Kao, 2000; Damon and Kristiansen, 2014). Third, the household has fewer members leading to lower consumption requirements but also less family labor available. The overall influence of these changes on the food consumption and nutrition of the household can be positive or negative.

Coming to the empirical studies analyzing the relationship between incidence of migration (both internal and international) and impact on household level outcomes, there are a mixed set of results. Hagen-Zanker and Azzari (2010), in the context of Albania, demonstrate that when the entire family moves from rural to peri-urban areas there is no change in consumption expenditure of migrant households. Karamba et al. (2011), in the case of Ghana, find that internal migration does not affect the total consumption expenditure of the migrant sending household but does affect the consumption pattern with a shift towards less nutritious food and eating out of home. Azzarri and Zezza (2011), for Tajikistan, find that international migration positively affects the outcomes of children left behind in terms of height for age (z-score) and incidence of stunting. Nguyen and Winters (2011), analysed the impact of short-term and long-term internal migration on food consumption pattern of households in

Vietnam. They document that both types of migration positively increase the food expenditure and ensures food security in short and long-term period. Using a unique migration survey from Tonga where migrants are decided based on a random lottery, Gibson et al. (2011) show that children left behind lag in health outcomes whereas children who accompany improve in the outcomes. This leads to a divergence in the outcomes of children. Beegle et al. (2011) study for Tanzania reveals an increase in consumption expenditure of international migrant households in comparison with households without migrants. In the context of India, Chandrasekhar et al. (2015) observe that households with short-term internal migrants have lower monthly per capita consumption and food expenditure in contrast to households without. This occurred as the short-term migrants were mostly employed in the unorganized sector which gave a low salary and lacked job security.

These strands highlight that important migration as a channel remains in improving the livelihood and food security of left-behind households. Further, they highlight that the final effect of migration on the outcomes of left-behind households may not always be positive and can also have adverse effects in some cases.

In this study, we investigate the impact of international migration on the food and nutritional outcomes of the left-behind household at origin using survey data from Kerala Migration Survey, 2011. The outcomes of interest are overall consumption expenditure, expenditure on food, share of food expenditure as well as dietary diversity measures- Shannon and Simpson Index. Our key findings are as follows. First, international migration by the member of the household increases the overall consumption expenditure as well as expenditure on food. Second, the effect of international migration on food consumption and dietary diversity in urban areas is of larger magnitude than their rural counterparts. Third, households with international migrants have higher dietary diversity than non-migrants. However, this result hides the heterogeneity in the food sub-groups consumption. The emigrant households spend more on protein intake and as well as on fruits and vegetables. Further, there is also increase in consumption of processed and ready to eat food items. This is a novel finding. Lastly, we also observe reduction in consumption of tobacco, liquor and intoxicants for the rural emigrant households.

The rest of the paper is organized as follows. The next section discusses the context of our study. Section 3 explains the data used. Section 4 provides the empirical modeling and estimation strategy. Section 5 provides the main results. Section 6 concludes with the discussion.

2. The context

The study focuses on the state of Kerala in India. Kerala is looked upon as the model for social development by most states in India, due to its remarkable performance in social indicators like high literacy rate, low infant mortality rate and birth rate (Rammohan, 2000). But this was not the scenario during the 1970s, the economy was in a miserable state with widespread poverty and unemployment. The average per capita expenditure was below national average and the two basic sectors of economy (agriculture and industry) were stagnant (Prakash, 1988; Kannan and Hari, 2002; Harilal and Joseph, 2003). This led the households in Kerala to diversify their source of livelihood, and thus they started moving within and outside India (Nair and Rama kumar, 2007). The reason for migration has mostly been revolving around the push and pull factors. The push factors include a high density of population, declining workforce participation rate, unemployment rate (among educated and uneducated), stagnant agriculture and backward industrial sector and the pull factors are better employment and earning opportunities. The most crucial among these are lack of adequate employment opportunities, a mismatch between the education profiles and available opportunities and wage differential between home and another state/country which have impelled the Keralites to leave their homes, looking for better lives for themselves and their household. The prime destination for international migrants is the Gulf and most researchers have found that "nine out of ten migrants from Kerala can be found in the Gulf" (Rajan and Zachariah, 2018).

There has been a gradual increase in the number of international migrants from Kerala per 100 households from 21.4 in 1998 to 29 in 2011. Internal migrants per 100 households have remained largely stagnant from 10.9 in 1998 to 11.1 in 2011 (Zachariah and Rajan, 2015). At present, –international migrants from Kerala are found in the countries of the Arabian Peninsula and internal migrants in Indian states of Tamil Nadu, Karnataka, Maharashtra, New Delhi, Gujarat, Nagaland, and Manipur.

Migration, especially international migration, has also made a noteworthy impact on the state of the economy through a healthy influx of remittances. In 2011 alone, Kerala received INR 497 billion, which comprised of 31.2 percent of the Net State Domestic Product (NSDP) of Kerala (Zachariah and Rajan,2015). A rise in remittances has brought about an increase in per capita income of the left behind households. There has been a remarkable improvement in the quality and standard of living of migrant households with an increase or a change in the level of household expenditure and investment patterns (Zachariah and Rajan, 2016). This has caused migration to play a significant role in poverty alleviation and ease the unemployment problem in the state to a considerable extent thereby giving a push to Kerala's economy (Zachariah et al., 1999).

Having understood the background of the international mobility from Kerala, we set out to contribute to this literature by analysing the impact of international migration on the left-behind household.

3. Data and descriptive statistics

The analysis has been performed using a novel data set from Kerala Migration Survey (KMS) conducted in 2011 (December 2010- May 2011) (Kerala Migration Survey, 2011). This survey was carried out under a collaborative venture by the migration research unit of the Centre for Development Studies (CDS), Thiruvananthapuram and the Government of Kerala, and funded by the Department of Non-Resident Keralite Affairs (NORKA).

Covering a sample of 15,000 households it collects data on 11,700 households residing in rural area and remaining in urban area. A total of 65,044 individuals (50,950 rural 14,094 urban individuals) were surveyed.

For this survey, household (HH) is the ultimate sample unit. And KMS defines *a household as "a group of persons normally living together and taking food from the same kitchen'. Persons living outside for study or work are not normal members of the household. Migrants living in other countries (emigrants –EMig) and other states (out-migrants OMig) were not enumerated. Paying guests and living-in servants and non-relatives are also included"*. These households have been chosen by stratified multistage random sampling. Overall 15,000 HHs were interviewed from the 14 districts of Kerala. From each district, a minimum of 1000 households was chosen, and the remaining were distributed among the larger districts[1]. Dependent on the status of the members, the households are categorized as emigrant, out-migrant or non-migrant. An emigrant or out- migrant household is the one that reports "*as any person (at least one or more) who was a usual resident of this household, migrated out of Kerala and is still living outside Kerala for at least a year?*".

Additionally, the respondents were also probed about their present place of residence (within or outside India). "*Those who were living outside India at the time of the survey were*

---

[1] In each district there is a rural and urban area and it forms the strata. In Kerala, as there are 14 districts, thus 28 strata-14 each in the rural and urban areas. The number of sample households in a district is distributed between the districts rural and urban strata in proportion to the number of rural and urban households in the district according to the 2001 census (as data for census 2011 was unavailable during this survey). Some modifications were however made in some districts where urbanization was relatively rapid.

*identified as emigrants and those who were living in India but outside Kerala as out-migrants*." From this point onwards in this study, international migrants as referred as emigrants. Apart from socio-economic and demographic characteristics, information related to consumption expenditure, health and education, savings and investment behavior of the left-behind members and non-migrant households is also collected.

It is estimated that of the 15,000 households interviewed 35 percent of households have at least one migrant. Out of total households, 18 percent (2736) have at least one emigrant. The data from the survey indicates that emigration from Kerala is male dominant (above 75 percent) and most of them are unmarried. The typical age of an emigrants is between 25-34 years. The emigrants are largely Muslims. Most of the migrants at the time of migration were either self-employed, unemployed or were out of labor force.

This is the only survey which provides such detailed information on household's consumption expenditure of frequently consumed food items in the last one month and non-food items in the last one year prior to the survey. In order to compute total consumption expenditure of a household, we convert the values of non-food items for which information is obtainable for a 365 day period to a 30 day period expenditure (i.e. monthly expenditure) by dividing with 365 and multiplying by 30. Consequently, the sum is further added to spending on items measured for a 30 day period divided by the household size, to calculate the monthly per capita consumption expenditure (MPCE). Monthly per capita expenditure on food (MPCE food) is calculated by adding the expenditure on food items consumed in the last 30 days and dividing it by the household size. The share on food expenditure is arrived at by dividing MPCE food by MPCE.

Coming to the descriptive statistics, the emigrant households have a higher average and median MPCE in comparison non-migrant households. A similar scenario holds true for average and median MPCE for households with emigrants in a rural and urban area of Kerala (See table 1).

[Insert Table 1 Here]

The average and median monthly consumption expenditure on food (MPCE food) are higher for households with emigrants in comparison to households without migrants. The emigrants and have a higher MPCE and MPCE food based on their source of income. Majority of the emigrants (64 percent) are employed in the private sector; whereas for the non-migrants, their source of income varies from agricultural labor (7 percent) to non-agricultural sector (11 percent) and self-employed (7 percent). The share of food expenditure portrays that households without migrants (52 percent) and emigrants (49 percent) have a similar share of food expenditure (See Table 1).

In our study, apart from MPCE, we also calculate dietary diversity indices as an indicator of food security for the households.

### 3.1. Measurement of food security through dietary diversity

We use two of the widely used indicators for measuring dietary diversity, that is, Shannon and Simpson index. These indices are used to measure the variation in the consumption of food and are considered as indirect measures of nutritional outcomes in the absence of direct measures. Ruel (2003) argues that these indicators indirectly capture both input and outcome indicator of food security, providing a complete picture of the well-being of a household. Recent studies by Nguyen and Winters (2011) and Sharma and Chandrasekhar (2016) have also used these measures to measure the dietary diversity in household food consumption.

The Simpson index is measured as $1 - \sum_{i=1}^{n} w_i^2$ where $w_i$ refers to the share of expenditure on different food sub-groups; i = 1; 2; . . . n refers to the groupings of food sub-groups. The index value for Simpson lies between zero and one. As the value of the index increase, we say that household has more diversified food consumption pattern.

Correspondingly the Shannon index is measured as $-\sum_{i=1}^{n} w_i \ln(w_i)$ where $w_i$ holds the same definition as in Simpson index, just the range of the index varies between zero and ln (n). This index provides lower weights to food sub-groups with a higher share of food expenditure and vice versa. The Simpson index is more weighted on dominant food sub-groups in comparison to Shannon index.

Henceforth to comprehend the degree of association between the two indices they are standardized with mean being zero and standard deviation being one. We find a correlation coefficient of 0.93 between these two indicators. Table 1 depicts the average and median values for Simpson and Shannon Index in both rural and urban areas. Based on these numbers, we observe that dietary diversity is lower among the emigrant households as compared to non-migrant households.

In the next section, we lay out the empirical model and estimation strategies used to analyse the relationship between incidence of emigration and food security of left-behind households.

4. Empirical model and estimation strategy

In this study, we estimate the impact of international migration (emigration) on the food security of left-behind household using monthly per capita consumption expenditure (log of food expenditure, share of food expenditure) and dietary diversity indices.

First, we estimate an ordinary least squares (OLS) regression model at the household level,

$$Y_i = \beta_0 + \beta_1 EMig_i + \beta_3 X_i + \varepsilon_i \quad (1)$$

Here $Y_i$ denotes the food security indicators and $EMig_i$ is the dummy variable for households with at least one emigrant. $X_i$ are explanatory variables which represent the socio-economic characteristics of household i and the characteristics of the head of the household i. The model includes the following additional explanatory variables ($X_i$), size of the household (1-2,3-5,6-10, and above 10 persons); whether household has a ration card; religion (Hindu, Christian, Muslim); type of the household occupying (luxurious, very good, poor, kutcha)[2] ; and the household head characteristics - gender; marital status (unmarried, married or widow/divorce); age (age group 0-14 years; 15-24 years; 25-34 years; 35-44 years; 45-59 years and 60 years and above); education level (primary, upper primary, secondary, higher secondary, degree and higher education, illiterate and literate without school education) and economic activity (employed by state/central/semi government aided, employed in private sector, self-employed and others, unemployed and economically inactive group). The size of land possessed (in acres) is a dummy variable coded as none, 0.01-0.100, 0.101-0.200, 0.201-0.300, 0.301-0.500, 0.501-1.00, 1.01-3.00 and more than 3.01 acres. The summary statistics for sampled households is given in Table 2.

[Insert Table 2 Here]

One major problem with this OLS model with the cross-section data is the endogeneity of the explanatory variables ($EMig_i$ in our case). This endogeneity can be on account of omitted variable bias and simultaneity bias in the model. The first bias refers to a situation where some important explanatory variables are missing from the model which can be correlated with the explanatory variables as well as the outcome variables. For example, local labor market conditions at the origin can affect the earnings and consumption pattern of the household as well as can be related to the incidence of migration. If such variables are not included in the model, the estimated coefficient of migration variables is likely to be biased. The second bias refers to the fact that level of consumption well-being of households remains an important input in the migration decision and migration affects the consumption pattern of the household. Therefore, any estimation using cross-section data and OLS approach is likely to be biased and have inconsistent results.

---

[2]As per KMS -2011 (i) "luxurious house" refers to 3 or more bedrooms with attached bathrooms, concrete roof, mosaic floor; (ii) "very good house" indicates -2 bedrooms with attached bathrooms, concrete roof, mosaic floor; (iii) "good house" implies one bedroom, brick and cement walls, concrete or tile roof; (iv) "poor house" refers to brick walls, cement floor, tin or asbestos roof; and (v) and "kutcha house" indicates, mud walls, mud floor, and thatched roof.

To address this issue of endogeneity, we use instrumental variables (IV) approach. Studies in the relevant literature, Sasin and McKenzie (2007); Iversen et al. (2009); Karamba et al. (2011); Nguyen and Winters (2011) and Chandrasekhar et al. (2015) suggest that a good instrumental variable is the one that is correlated with the explanatory variable (i.e., the incidence of international migration) but uncorrelated with the outcome variable (i.e., indicators of consumption expenditure and diet diversity measures). This helps in eliminating the biases that arise due to endogeneity. Thus, we look for instruments that affect international migration and exogenously/indirectly affect household consumption expenditure, the share of food and diet diversity indices. Seeking help through existing literature on networks as an instrument for migration we could figure out certainly reasonable instruments for emigration. Studies in the past by McKenzie and Rapoport (2007), Acosta (2011) and Mendola and Carletto (2012) indicate networks act as an instrument for migration as it provides information to the migrants directly or indirectly on job opportunities. This was helpful in figuring out two relevant instruments for emigration from the survey itself.

The instrument variables used in our analysis are a) if offered work visa, will you be willing to emigrate (yes-1, no-0); b) whether know any relative, friend or neighbor who has emigrated outside India to work (yes -1, no-0). The explanation for their usage is as follows.

First, emigration is an expensive affair, with a huge cost being incurred in availing the VISA and purchasing of the ticket. The other costs incurred during the process also includes the cost borne by the recruitment agency; emigration clearance and passport which are meagre in comparison. The process of emigration involves a considerable amount of financial sacrifice especially if the emigrants are from poor families. Thus only a few Kerala people could emigrate without any financial support. If visa to work outside India is offered to any working-age member of the family, this would instigate/prompt him /her to emigrate. Therefore, first instrument identified to deal with the issue of endogeneity is "if offered a work visa" and KMS 2011 collects information on it. In the survey a detailed question confining to this aspect was

*if someone offered a work-visa to work outside India to any working-age member of your household, would he/she accept it and emigrate.*

Second, the role of migrant networks (direct and indirect) is well established in explaining the decision to migrate (Iversen et al., 2009; Karamba et al., 2011; Nguyen and Winters, 2011 and Chandrasekhar et al., 2015). Existing migrants help in reducing the search cost and cost of assimilation is lower for migrants with social ties at the destination. In our survey, a question relating to social network for migrants is asked: *"known any relative, friend or neighbor who has emigrated outside India to work*." This variable captures the existence of migration network within the family/friend/ neighborhood and is assumed to directly affect the migrant status of a household. Thus, a household with a known relative/friend outside India to work is identified as the second instrument to tackle the endogeneity issue.

The impact of international migration can be estimated using two-stage least square (2SLS) instrumental variable approach. It needs to be noted that the dummy variable has been created for both the instruments. The first stage estimates the relationship between the instruments as an explanatory variable and being an emigrant household as the dependent variable. And the dependent variable ($EMig_i$) is not correlated with the error term. The first stage is represented mathematically below:

$$EMig_i = \alpha_0 + \alpha_1 X_i + \alpha_2 MigNetwork_i + \alpha_3 WkVisa_i + \mu_i \qquad (2) \qquad [\text{Stage 1}]$$

where $EMig_i$ – the dependent variable is a dummy variable taking the value one or zero subject to whether or not the household i has an emigrant; $X_i$ is the explanatory variable comprising of socio economic characteristics of the household and its head ( as used in the OLS equation); $RAbroad_i$ and $WkVisa_i$ are the instruments; $MigNetwork_i$ is a dummy variable taking the value one or zero subject to whether or not the household i has a network of a relative /friend who has emigrated abroad to work and $WkVisa_i$ is a dummy variable taking the value one or zero subject to whether or not the work visa is provided/offered to the working-age member of household i.

In the second stage, the predicted value of the dependent variable ($EMig\_hat_i$) obtained in the first stage is used as an explanatory variable with other covariates ($X_i$) and error term ($\varepsilon_i$) to predict the relationship between emigration and indicators of consumption expenditure and dietary diversity indices. The second stage of the instrumental variable regression is expressed below:

$$Y_i = \beta_0 + \beta_1 EMig\_hat_i + \beta_2 X_i + \varepsilon_i \qquad (3) \qquad [\text{Stage 2}]$$

where $EMig_{ihat}$ is the predicted value of from stage 1 (equation 2); $Y_i$ is the dependent variable and $\varepsilon_i$ is the error term.

In the next section, we present the results of the study.

## 5. Estimation results

We discuss the results in two parts. First, we report the ordinary least square (OLS) results for rural and urban households. Next, we come to the instrumental variable estimates for the impact of international migration.

### 5.1. OLS estimates

Observing the estimates in table 3, panel A, we find that households with at least one emigrant (HHEM hereafter), has 23 percent higher consumption expenditure than households with no emigrants. Also, food expenditure is 13 percent higher in households with an emigrant. As is expected, the share of food expenditure is lower for emigrant households. These findings are in line with Nguyen and Winters (2011), Karamba et al. (2011) and Sharma and Chandrasekhar (2016). Coming to the Simpson and Shannon Index, we find that emigrant households have higher dietary diversity than households with no emigrant. In table 3, panel B and C, we highlight the results for rural and urban households separately. We find that there are significant differences in consumption expenditure and expenditure on food between rural and urban areas. However, we observe that dietary diversity is higher in households with emigrants in rural areas as compared to urban areas. These findings are in contrast to Nguyen and Winters (2011) who do not find any evidence of the effect of migration on dietary diversity.

[Insert Table 3 Here]

### 5.2. Instrumental Variable (IV) estimates

In this sub-section, we discuss the findings from the instrumental variable model. Before discussing the impact of international migration on expenditure and dietary diversity, we provide the diagnostic tests for the instrumental variable model.

First, we test whether the variable "Households with at least one emigrant" is endogenous or not. The tests for endogeneity of HHEM variable rejects the null hypothesis that this variable can be treated as exogenous in the estimating equation. The under-identification test suggests that the model is identified for the sample, and the Kleibergen–Paap rk LM statistic does not accept the null hypothesis that the model is under-identified. This test indicates that the excluded instruments are relevant, which means correlated with the endogenous HHEM variable. The overidentification test is satisfied indicating that the instruments are jointly significant for the model. In our IV models, the explanatory power of the instruments used is quite strong, with F-statistics having a reasonably high value (i.e., above 10).

[Insert Table 4 Here]

Coming to the main results from the instrumental variable models, in table 5, we show that households with at least one emigrant (HHEM) have higher MPCE, higher food expenditure and share of food expenditure reduces. Also, we find that, based on both Shannon and Simpson index, HHEM have higher dietary diversity than non-emigrant households. Further, in comparison to OLS results (see table 3), we observe that IV estimates are larger in magnitude indicating the downward bias on OLS estimates. Further, comparing the rural and urban samples, we document that there are no significant differences in terms of MPCE and food expenditure (MPCE food) and share of food expenditure. However, for dietary diversity measures (Shannon and Simpson indices), we show that urban households with at least one emigrants consume a more diverse diet as compared to their rural counterparts. One plausible explanation for the same is as follows. It could be possible that the remittances and information channel for both rural and urban households might be equally active. However, urban households can have access to more diverse set of food items which is translating the information content of migration into actual consuming a more diverse diet.

[Insert Table 5 Here]

A natural question that arises from this discussion is what sort of food items are being consumed by the urban and rural emigrant households in their diverse diet. To answer this question, we analyse the differences in spending on food sub-groups among emigrant and non-emigrant households in rural and urban areas. In table 6, we report the results for OLS and IV models for rural, urban and combined (rural + urban) households. Some key observations in these results are as follows. There is higher spending on protein (pulses and milk, milk products) in both rural and urban emigrant households. Rural emigrant households spend more on vegetables and fruits, egg, fish and meat, but urban emigrant households do not. On the other hand, urban and rural emigrant households spend more on processed and ready to eat food items, but the magnitude is higher for the urban households. This validates our earlier plausible explanation. Further, this finding validates that though dietary diversity is increasing among urban and rural household with at least one emigrant, there is a negative impact in terms of spending on processed and ready to food items which are considered to be less healthy options. We can argue that information channel of migration, which exposes the left-behind households with various type of food habits, can lead to negative impact on health and nutrition, if it promotes the consumption of less healthy food items (processed and ready to eat food).

[Insert Table 6 Here]

Last but not the least, we also analyse the impact of emigration on consumption of tobacco, liquor and intoxicants. We show that for rural emigrant households, the consumption of these items reduces whereas no such effect is present for urban emigrant households. One plausible explanation could be that the household member who used to spend on these items, has migrated leading to lower spending on tobacco, liquor and intoxicants. However, the similar effect for urban households is not present.

In the next, section, we provide the summary of the paper, with concluding remarks.

6. Conclusion

In this study, we investigate the impact of international migration on the consumption expenditure, food expenditure, and dietary diversity of the left-behind households in the context of India. We use the Kerala Migration Survey and the quasi-experimental method of instrumental variable approach to answer his question. The main findings from our study are as follows. First, the overall consumption and food expenditure of the emigrant households improves due to emigration of at least one household member. Second, such migration leads to higher dietary diversity for both rural and urban left-behind households. However, this result hides the heterogeneity in the food sub-groups consumption. Third, the emigrant households spend more on protein intake and as well as on fruits and vegetables. Further, there is also increase in consumption of processed and ready to eat food items. This is a novel finding. Lastly, we also observe reduction in consumption of tobacco, liquor and intoxicants for the rural emigrant households.

Some of the key implications of these findings are as follows for various stakeholders ranging from household to policymakers. First, from the household's perspective, we show that international migration improves the overall well-being of the left-behind household in terms of consumption expenditure and dietary diversity. However, some cautions are important to highlight. International migration may not always have just beneficial impact. As seen in our study, international migration by the members exposes the household to unhealthy food habits such as consumption of processed and ready to eat food items. For the policymakers, our findings are useful, as they highlight how there could be differential impact of international migration on households in rural and urban areas. It is important to know the impact on consumption to understand the changing nutritional and health outcomes of left-behind household members. Policymakers can use the findings of this study to understand the nature of interventions to amplify the positive effects of migration while at the same time neutralizing the negative effects (if any).

TABLES

Table 1: Descriptive Statistics for various types of households by migration status

| | | Type of households | | | |
|---|---|---|---|---|---|
| | | Emigrants | | Non-migrants | |
| | Variables | Mean | Median | Mean | Median |
| Rural | MPCE (in rupees) | 1922.75 | 1493.75 | 1484.06 | 1265 |
| | MPCE food expenditure (in rupees) | 822.86 | 752 | 707.51 | 652.25 |
| | Share of food expenditure (%) | 49.77 | 51.24 | 52.61 | 53.16 |
| | Dietary Diversity Measures | | | | |
| | Simpson Index (standardized) | -0.0038 | 0.0995 | 0.004 | 0.128 |
| | Shannon Index (standardized) | -0.0133 | -0.045 | 0.0157 | 0.0055 |
| Urban | MPCE (in rupees) | 2376.31 | 1771.35 | 1729.81 | 1468.33 |
| | MPCE food expenditure (in rupees) | 941.21 | 836.63 | 797.56 | 746.25 |
| | Share of food expenditure (%) | 47.46 | 48.35 | 51.38 | 52.02 |
| | Dietary Diversity Measures | | | | |
| | Simpson Index (standardized) | 0.022 | 0.042 | 0.18 | 0.127 |
| | Shannon Index (standardized) | 0.0031 | 0.0044 | 0.0164 | 0.031 |
| Total | MPCE (in rupees) | 2024.2 | 1542.55 | 1538.58 | 1311.75 |
| | MPCE food expenditure (in rupees) | 849.33 | 768.29 | 727.49 | 675.33 |
| | Share of food expenditure (%) | 49.25 | 50.59 | 52.34 | 52.87 |
| | Dietary Diversity Measures | | | | |
| | Simpson Index (standardized) | 0.003 | 0.107 | 0.008 | 0.129 |
| | Shannon Index (standardized) | -0.008 | -0.015 | 0.016 | 0.009 |

Source: Author's calculation using micro data from Kerala Migration Survey, 2011

| Table 2: Summary Statistics for sampled households | | | | |
|---|---|---|---|---|
| Variables | Mean | Standard Deviation | Min. | Max. |
| Emigrant Household | 0.1824 | - | 0 | 1 |
| Logarithm (MPCE) | 7.2712 | 0.545 | 5.575 | 10.887 |
| Logarithm (MPCE food expenditure) | 6.551 | 0.409 | 4.933 | 8.702 |
| Share of food expenditure | 0.509 | 0.135 | 0.317 | 0.939 |
| Dietary Diversity Measures | | | | |
| Simpson Index | 0.865 | 0.032 | 0.585 | 0.931 |
| Shannon Index | 2.271 | 0.167 | 1.418 | 2.698 |
| Simpson Index (standardized) | 0 | 1 | -8.79 | 2.066 |
| Shannon Index (standardized) | 0 | 1 | -5.11 | 2.561 |
| Dependency Ratio | 52.92 | 64.51 | 0 | 500 |
| Gender of the Household Head | | | | |
| Female | 0.26 | - | 0 | 1 |
| Male | 0.74 | - | 0 | 1 |
| Age Group of the Household Head (in years) | | | | |
| 0-14 | 0.0001 | - | - | -- |
| 15-24 | 0.002 | - | 0 | 1 |
| 25-34 | 0.032 | - | 0 | 1 |
| 35-44 | 0.1495 | - | 0 | 1 |
| 45-59 | 0.3981 | - | 0 | 1 |
| 60+ | 0.4183 | - | 0 | 1 |
| Marital Status of Household Head | | | | |
| Unmarried | 0.011 | - | 0 | 1 |
| Married | 0.763 | - | 0 | 1 |
| Divorced and Others | 0.225 | - | 0 | 1 |
| Religion | | | | |
| Hindu | 0.592 | - | 0 | 1 |
| Christian | 0.198 | - | 0 | 1 |
| Muslim | 0.21 | - | 0 | 1 |
| Educational Level of household head | | | | |
| Primary | 0.206 | | 0 | 1 |
| Upper Primary | 0.214 | - | 0 | 1 |
| Secondary | 0.353 | - | 0 | 1 |
| Higher Secondary | 0.046 | - | 0 | 1 |
| Degree and Higher Education | 0.078 | - | 0 | 1 |
| illiterate | 0.087 | - | 0 | 1 |
| Literate without school education | 0.017 | - | 0 | 1 |
| Economic Activity | | | | |
| Employed by State/Central/Semi Govt Aided | 0.027 | | 0 | 1 |
| Employed in Private Sector | 0.066 | - | 0 | 1 |
| Self Employed and Others | 0.486 | - | 0 | 1 |
| Unemployed | 0.003 | - | 0 | 1 |
| Economically Inactive Group | 0.417 | - | 0 | 1 |
| Size of the Household (no. of persons) | | | | |
| 1-2 | 0.155 | | 0 | 1 |
| 3-5 | 0.635 | - | 0 | 1 |
| 6-10 | 0.197 | - | 0 | 1 |
| 10+ | 0.013 | - | 0 | 1 |
| Land Holding Size (in acres) | | - | | |
| None | 0.719 | - | 0 | 1 |
| 0.01-0.100 | 0.07 | - | 0 | 1 |
| 0.101-0.200 | 0.038 | - | 0 | 1 |
| 0.201-0.300 | 0.029 | - | 0 | 1 |
| 0.301-0.500 | 0.044 | - | 0 | 1 |
| 0.501-1.00 | 0.051 | - | 0 | 1 |
| 1.01-3.00 | 0.037 | - | 0 | 1 |
| More than 3.01 | 0.011 | - | 0 | 1 |
| Observations | 15000 | | | |

| | (1) | (2) | (3) | (4) | (5) |
|---|---|---|---|---|---|
| | | | Share of | Simpson | Shannon |
| VARIABLES | Ln(MPCE) | Ln(MPCE food) | MPCE food | Index | Index |
| **Panel A: Overall (rural +urban)** | | | | | |
| Household with no emigrant (base cat.) | | | | | |
| Household with an emigrant | 0.235*** | 0.136*** | -0.0418*** | 0.135*** | 0.139*** |
| | (0.0101) | (0.00741) | (0.00292) | (0.0220) | (0.0206) |
| Observations | 12,430 | 12,430 | 12,430 | 12,430 | 12,430 |
| R-squared | 0.431 | 0.488 | 0.235 | 0.277 | 0.362 |
| **Panel B: Rural** | | | | | |
| Household with no emigrant (base cat.) | | | | | |
| Household with an emigrant | 0.233*** | 0.138*** | -0.0420*** | 0.145*** | 0.162*** |
| | (0.0114) | (0.00831) | (0.00330) | (0.0252) | (0.0231) |
| Observations | 9,668 | 9,668 | 9,668 | 9,668 | 9,668 |
| R-squared | 0.436 | 0.497 | 0.257 | 0.277 | 0.386 |
| **Panel C: Urban** | | | | | |
| Household with no emigrant (base cat.) | | | | | |
| Household with an emigrant | 0.246*** | 0.135*** | -0.0431*** | 0.136*** | 0.110*** |
| | (0.0208) | (0.0150) | (0.00601) | (0.0432) | (0.0423) |
| Observations | 2,762 | 2,762 | 2,762 | 2,762 | 2,762 |
| R-squared | 0.473 | 0.529 | 0.247 | 0.351 | 0.396 |

Table 3: OLS estimates for left-behind households

Standard errors in parentheses, *** p<0.01, ** p<0.05, * p<0.1
Only select variables of interest are reported. For detailed result tables refer to appendix table A1, A2, A3.

| Table 4: Diagnostic tests for Instrumental variable models | |
|---|---|
| Test | Household with an emigrant |
| Endogeneity test | 30.874 (0.00) |
| Weak Instrument test (Sanderson-Windmeijer (SW) test) | 114.35 (0.00) |
| Under identification test (Kleibergen- Papp rk LM statistics) | 132.94 (0.00) |
| Weak identification test (Cragg-Donald Wald F statistics) | 621.71 (0.00) |
| Hansen J Statistics | 0.725 (0.39) |

p-value are in parenthesis

| | (1) | (2) | (3) | (4) | (5) |
|---|---|---|---|---|---|
| | | | Share of MPCE | Simpson | Shannon |
| VARIABLES | ln(MPCE) | ln(MPCE food) | food | Index | Index |
| **Panel A: Overall (rural +urban)** | | | | | |
| Household with no emigrant (base cat.) | | | | | |
| Household with an emigrant | 0.594*** | 0.295*** | -0.130*** | 0.625*** | 0.755*** |
| | (0.0697) | (0.0577) | (0.0179) | (0.174) | (0.174) |
| Observations | 12,431 | 12,431 | 12,431 | 12,431 | 12,431 |
| R-squared | 0.373 | 0.469 | 0.178 | 0.247 | 0.316 |
| **Panel B: Rural** | | | | | |
| Household with no emigrant (base cat.) | | | | | |
| Household with an emigrant | 0.582*** | 0.284*** | -0.133*** | 0.472** | 0.570*** |
| | (0.0772) | (0.0643) | (0.0196) | (0.187) | (0.179) |
| Observations | 9,668 | 9,668 | 9,668 | 9,668 | 9,668 |
| R-squared | 0.381 | 0.480 | 0.199 | 0.264 | 0.366 |
| **Panel C: Urban** | | | | | |
| Household with no emigrant (base cat.) | | | | | |
| Household with an emigrant | 0.567*** | 0.244** | -0.131*** | 0.941** | 1.095** |
| | (0.137) | (0.115) | (0.0349) | (0.374) | (0.454) |
| Observations | 2,763 | 2,763 | 2,763 | 2,763 | 2,763 |
| R-squared | 0.427 | 0.520 | 0.188 | 0.268 | 0.275 |

Table 5: Instrumental Variable estimates for left-behind households

Standard errors in parentheses, *** p<0.01, ** p<0.05, * p<0.1

Only select variables of interest are reported. For detailed result tables refer to appendix tables A4, A5, A6.

| | OLS estimates | | | IV estimates | | |
|---|---|---|---|---|---|---|
| Dependent variables | Overall (rural+ urban) | Rural | Urban | Overall (rural+ urban) | Rural | Urban |
| | (1) | (2) | (3) | (4) | (5) | (6) |
| MPCE spendings on various sub-groups | (Explanatory variable: household with an emigrant) | | | | | |
| Cereals | 0.119*** | 0.115*** | 0.0575*** | 0.248*** | 0.323*** | 0.0366 |
| | (0.0103) | (0.0124) | (0.0211) | (0.0790) | (0.0853) | (0.181) |
| Pulses | 0.0606*** | 0.0579*** | 0.0120 | 0.429*** | 0.472*** | 0.386** |
| | (0.0101) | (0.0104) | (0.0224) | (0.0919) | (0.103) | (0.182) |
| Milk and related products | 0.201*** | 0.179*** | 0.213*** | 0.526*** | 0.505*** | 0.544*** |
| | (0.0146) | (0.0168) | (0.0302) | (0.0974) | (0.108) | (0.172) |
| Oil, Ghee and butter etc. | 0.110*** | 0.0917*** | 0.102*** | 0.116 | 0.106 | 0.257 |
| | (0.0107) | (0.0122) | (0.0220) | (0.0886) | (0.0914) | (0.240) |
| Vegetables and fruits | 0.173*** | 0.177*** | 0.0957*** | 0.314*** | 0.385*** | 0.0690 |
| | (0.0127) | (0.0149) | (0.0260) | (0.0822) | (0.0937) | (0.175) |
| Egg, fish and meat | 0.218*** | 0.233*** | 0.143** | 0.178 | 0.274* | 0.0810 |
| | (0.0239) | (0.0309) | (0.0680) | (0.126) | (0.150) | (0.339) |
| Sugar | 0.0674*** | 0.0590*** | 0.0279 | 0.162* | 0.235** | -0.105 |
| | (0.0105) | (0.0119) | (0.0232) | (0.0841) | (0.0926) | (0.200) |
| Processed food items | 0.318*** | 0.359*** | 0.443*** | 0.864*** | 0.944*** | 1.656** |
| | (0.0312) | (0.0453) | (0.0859) | (0.207) | (0.265) | (0.742) |
| Ready to eat food | -0.0157 | -0.0199 | -0.0294 | 1.251*** | 1.216*** | 3.234*** |
| | (0.0382) | (0.0564) | (0.112) | (0.284) | (0.361) | (1.217) |
| Tobacco, liquor and intoxicants | -0.224*** | -0.329*** | -0.333** | -0.217 | -0.745** | 1.038 |
| | (0.0488) | (0.0729) | (0.130) | (0.259) | (0.357) | (0.856) |

Table 6: Sub-group wise expenditure estimates for left-behind households

Standard errors in parentheses, *** p<0.01, ** p<0.05, * p<0.1
Only select variables of interest are reported.
For detailed result tables refer to appendix A, tables A4, A5, A6, A7, A8, A9.

# APPENDIX

Table A1: OLS estimates for households left behind (rural +urban)

| VARIABLES | (1) Ln(MPCE) | (2) Ln(MPCE food) | (3) Share of food expenditure | (4) Simpson | (5) Shannon |
|---|---|---|---|---|---|
| household with no emigrant (base cat.) | | | | | |
| household with an emigrant | 0.235*** | 0.136*** | -0.0418*** | 0.135*** | 0.139*** |
| | (0.0101) | (0.00741) | (0.00292) | (0.0220) | (0.0206) |
| Dependency ratio | -0.000665*** | -0.000509*** | 5.37e-05*** | 0.000261* | 0.000372*** |
| | (6.32e-05) | (4.62e-05) | (1.82e-05) | (0.000137) | (0.000129) |
| Gender of household head (base cat: male) | | | | | |
| Female | -0.00432 | -0.0233** | -0.00827* | 0.00184 | -0.0211 |
| | (0.0148) | (0.0108) | (0.00427) | (0.0321) | (0.0302) |
| Age of household head (base cat: 15-24 years) | | | | | |
| 25-34 years | -0.0318 | 0.0213 | 0.0177 | 0.319* | 0.320* |
| | (0.0819) | (0.0599) | (0.0237) | (0.178) | (0.167) |
| 35-44 years | 0.00257 | 0.0122 | -0.00244 | 0.215 | 0.220 |
| | (0.0804) | (0.0588) | (0.0232) | (0.174) | (0.164) |
| 45-59 years | 0.0860 | 0.0453 | -0.0239 | 0.241 | 0.243 |
| | (0.0801) | (0.0586) | (0.0231) | (0.174) | (0.163) |
| 60 and above years | 0.0758 | 0.0412 | -0.0218 | 0.223 | 0.231 |
| | (0.0802) | (0.0587) | (0.0232) | (0.174) | (0.163) |
| Religion of household head (base cat: Hindu) | | | | | |
| Christian | 0.0908*** | 0.0790*** | -0.00400 | -0.0214 | -0.0108 |
| | (0.00994) | (0.00727) | (0.00287) | (0.0216) | (0.0203) |
| Muslim | 0.0138 | 0.0496*** | 0.0141*** | 0.00384 | 0.0411* |
| | (0.0107) | (0.00779) | (0.00308) | (0.0231) | (0.0217) |
| Education level of household head (base cat: Primary) | | | | | |
| Upper Primary | 0.0467*** | 0.0223*** | -0.0123*** | 0.0694*** | 0.0869*** |
| | (0.0112) | (0.00818) | (0.00323) | (0.0243) | (0.0228) |
| Secondary | 0.169*** | 0.0913*** | -0.0363*** | 0.0969*** | 0.127*** |
| | (0.0109) | (0.00798) | (0.00315) | (0.0237) | (0.0222) |
| Higher Secondary | 0.348*** | 0.172*** | -0.0745*** | 0.182*** | 0.230*** |
| | (0.0199) | (0.0145) | (0.00574) | (0.0431) | (0.0405) |
| Degree and above | 0.458*** | 0.259*** | -0.0870*** | 0.251*** | 0.344*** |
| | (0.0173) | (0.0126) | (0.00499) | (0.0375) | (0.0352) |
| not educated | -0.102*** | -0.0959*** | 0.00243 | -0.0919*** | -0.0836*** |
| | (0.0147) | (0.0108) | (0.00426) | (0.0320) | (0.0300) |
| Literate without school education | -0.0295 | -0.0131 | 0.00777 | 0.156** | 0.177*** |
| | (0.0280) | (0.0205) | (0.00809) | (0.0609) | (0.0571) |
| Marital status of household head (base cat: Unmarried) | | | | | |
| Married | -0.00597 | 0.0220 | 0.00812 | 0.163** | 0.145** |
| | (0.0345) | (0.0252) | (0.00996) | (0.0749) | (0.0703) |
| Divorced and others | -0.0365 | 0.00787 | 0.0156 | 0.199*** | 0.168** |
| | (0.0352) | (0.0257) | (0.0102) | (0.0764) | (0.0717) |
| Size of household (base cat: 0-2) | | | | | |
| 3 - 5 members | -0.279*** | -0.300*** | -0.0165*** | 0.00785 | 0.129*** |
| | (0.0107) | (0.00785) | (0.00310) | (0.0233) | (0.0219) |
| 6-10 members | -0.501*** | -0.492*** | -0.00374 | -0.380*** | -0.0975*** |
| | (0.0133) | (0.00973) | (0.00384) | (0.0289) | (0.0271) |
| 10 and above members | -0.725*** | -0.673*** | 0.0119 | -0.514*** | -0.0963 |
| | (0.0358) | (0.0262) | (0.0103) | (0.0777) | (0.0729) |
| Land holdings in acres (base cat: no land) | | | | | |
| 0.01-0.100 | 0.108*** | 0.0559*** | -0.0260*** | 0.0980*** | 0.103*** |
| | (0.0154) | (0.0113) | (0.00444) | (0.0334) | (0.0313) |
| 0.101-0..200 | 0.0928*** | 0.0692*** | -0.0122** | 0.105** | 0.101** |
| | (0.0197) | (0.0144) | (0.00570) | (0.0429) | (0.0402) |
| 0.201-0.300 | 0.0658*** | 0.0598*** | -0.00408 | 0.0427 | 0.0384 |
| | (0.0221) | (0.0162) | (0.00638) | (0.0480) | (0.0450) |
| 0.301-.0.500 | 0.0733*** | 0.0777*** | -0.00287 | 0.0578 | 0.0342 |
| | (0.0191) | (0.0140) | (0.00551) | (0.0414) | (0.0389) |
| 0.501-1.000 | 0.175*** | 0.120*** | -0.0281*** | 0.157*** | 0.125*** |
| | (0.0183) | (0.0134) | (0.00528) | (0.0397) | (0.0373) |
| 1.001-3.000 | 0.210*** | 0.123*** | -0.0392*** | 0.129*** | 0.120*** |
| | (0.0208) | (0.0152) | (0.00599) | (0.0451) | (0.0423) |
| More than 3.000 | 0.247*** | 0.156*** | -0.0409*** | 0.226*** | 0.198*** |
| | (0.0367) | (0.0269) | (0.0106) | (0.0797) | (0.0748) |
| Economic activity of household head (base cat: Govt. job) | | | | | |
| Private sector job | -0.0755*** | -0.0489** | 0.0144* | -0.0186 | -0.0137 |
| | (0.0261) | (0.0191) | (0.00753) | (0.0566) | (0.0532) |
| Self employed | -0.110*** | -0.0696*** | 0.0207*** | -0.0477 | -0.0329 |
| | (0.0236) | (0.0173) | (0.00683) | (0.0513) | (0.0482) |
| Unemployed | -0.0540 | -0.109* | -0.0210 | -0.00364 | -0.0741 |
| | (0.0798) | (0.0584) | (0.0230) | (0.173) | (0.163) |
| Economically inactive | -0.0523** | -0.0449** | 0.00578 | -0.00935 | 0.00224 |
| | (0.0246) | (0.0180) | (0.00709) | (0.0533) | (0.0501) |
| Constant | 7.437*** | 6.901*** | 0.617*** | -0.684*** | -0.985*** |
| | (0.0871) | (0.0637) | (0.0251) | (0.189) | (0.177) |
| | | | | | |
| Observations | 12,430 | 12,430 | 12,430 | 12,430 | 12,430 |
| R-squared | 0.431 | 0.488 | 0.235 | 0.277 | 0.362 |

Standard errors in parentheses
*** p<0.01, ** p<0.05, * p<0.1

| | (1) | (2) | (3) | (4) | (5) |
|---|---|---|---|---|---|
| VARIABLES | Ln(MPCE) | Ln(MPCE food) | Share of food expenditure | Simpson | Shannon |
| household with no emigrant (base cat.) | | | | | |
| household with an emigrant | 0.233*** | 0.138*** | -0.0420*** | 0.145*** | 0.162*** |
| | (0.0114) | (0.00831) | (0.00330) | (0.0252) | (0.0231) |
| Dependency ratio | -0.000679*** | -0.000519*** | 5.81e-05*** | 0.000275* | 0.000352** |
| | (6.98e-05) | (5.10e-05) | (2.03e-05) | (0.000155) | (0.000142) |
| Gender of household head (base cat: male) | | | | | |
| Female | -0.0110 | -0.0344*** | -0.0117** | -0.00749 | -0.0283 |
| | (0.0165) | (0.0121) | (0.00480) | (0.0366) | (0.0336) |
| Age of household head (base cat: 15-24 years) | | | | | |
| 25-34 years | -0.0171 | 0.0290 | 0.0126 | 0.196 | 0.131 |
| | (0.0940) | (0.0687) | (0.0273) | (0.208) | (0.191) |
| 35-44 years | 0.00838 | 0.0223 | -0.00262 | 0.0747 | 0.0151 |
| | (0.0925) | (0.0676) | (0.0269) | (0.205) | (0.188) |
| 45-59 years | 0.0780 | 0.0458 | -0.0229 | 0.0968 | 0.0328 |
| | (0.0922) | (0.0674) | (0.0268) | (0.204) | (0.188) |
| 60 and above years | 0.0527 | 0.0377 | -0.0153 | 0.0844 | 0.0252 |
| | (0.0925) | (0.0676) | (0.0269) | (0.205) | (0.188) |
| Religion of household head (base cat: Hindu) | | | | | |
| Christian | 0.109*** | 0.0864*** | -0.00899*** | -0.0452* | -0.0414* |
| | (0.0111) | (0.00813) | (0.00323) | (0.0247) | (0.0226) |
| Muslim | 0.0306** | 0.0673*** | 0.0154*** | -0.00862 | 0.0182 |
| | (0.0120) | (0.00877) | (0.00349) | (0.0266) | (0.0244) |
| Education level of household head (base cat: Primary) | | | | | |
| Upper Primary | 0.0516*** | 0.0253*** | -0.0133*** | 0.0486* | 0.0696*** |
| | (0.0122) | (0.00891) | (0.00354) | (0.0270) | (0.0248) |
| Secondary | 0.154*** | 0.0849*** | -0.0333*** | 0.0686** | 0.0966*** |
| | (0.0121) | (0.00887) | (0.00353) | (0.0269) | (0.0247) |
| Higher Secondary | 0.325*** | 0.165*** | -0.0668*** | 0.123** | 0.157*** |
| | (0.0221) | (0.0162) | (0.00642) | (0.0490) | (0.0450) |
| Degree and above | 0.411*** | 0.245*** | -0.0770*** | 0.165*** | 0.243*** |
| | (0.0213) | (0.0156) | (0.00618) | (0.0472) | (0.0433) |
| not educated | -0.0936*** | -0.0870*** | 0.00274 | -0.0860** | -0.0725** |
| | (0.0159) | (0.0116) | (0.00462) | (0.0352) | (0.0323) |
| Literate without school education | -0.0398 | -0.0142 | 0.0100 | 0.0673 | 0.0943 |
| | (0.0305) | (0.0223) | (0.00887) | (0.0677) | (0.0621) |
| Marital status of household head (base cat: Unmarried) | | | | | |
| Married | -0.0135 | 0.00489 | 0.00642 | 0.148* | 0.118 |
| | (0.0391) | (0.0286) | (0.0114) | (0.0867) | (0.0796) |
| Divorced and others | -0.0490 | -0.0160 | 0.0121 | 0.151* | 0.111 |
| | (0.0398) | (0.0291) | (0.0116) | (0.0883) | (0.0811) |
| Size of household (base cat: 0-2) | | | | | |
| 3 - 5 members | -0.269*** | -0.298*** | -0.0204*** | -0.0270 | 0.0928*** |
| | (0.0120) | (0.00875) | (0.00348) | (0.0265) | (0.0244) |
| 6-10 members | -0.489*** | -0.488*** | -0.00804* | -0.426*** | -0.148*** |
| | (0.0148) | (0.0108) | (0.00430) | (0.0328) | (0.0301) |
| 10 and above members | -0.711*** | -0.681*** | 0.000630 | -0.462*** | -0.0713 |
| | (0.0389) | (0.0284) | (0.0113) | (0.0861) | (0.0791) |
| Land holdings in acres (base cat: no land) | | | | | |
| 0.01-0.100 | 0.107*** | 0.0493*** | -0.0306*** | 0.0756** | 0.0873** |
| | (0.0173) | (0.0126) | (0.00503) | (0.0383) | (0.0352) |
| 0.101-0..200 | 0.0643*** | 0.0452*** | -0.0124* | 0.0860* | 0.0836* |
| | (0.0218) | (0.0159) | (0.00633) | (0.0483) | (0.0443) |
| 0.201-0.300 | 0.0634*** | 0.0497*** | -0.00657 | 0.0256 | 0.0155 |

|  | (0.0239) | (0.0175) | (0.00696) | (0.0530) | (0.0487) |
|---|---|---|---|---|---|
| 0.301-.0.500 | 0.0783*** | 0.0793*** | -0.00374 | 0.0862* | 0.0622 |
|  | (0.0202) | (0.0148) | (0.00588) | (0.0448) | (0.0412) |
| 0.501-1.000 | 0.187*** | 0.127*** | -0.0296*** | 0.184*** | 0.165*** |
|  | (0.0194) | (0.0142) | (0.00564) | (0.0430) | (0.0395) |
| 1.001-3.000 | 0.208*** | 0.118*** | -0.0395*** | 0.149*** | 0.142*** |
|  | (0.0215) | (0.0157) | (0.00625) | (0.0477) | (0.0438) |
| More than 3.000 | 0.251*** | 0.164*** | -0.0383*** | 0.248*** | 0.229*** |
|  | (0.0380) | (0.0277) | (0.0110) | (0.0841) | (0.0772) |
| Economic activity of household head (base cat: Govt. job) | | | | | |
| Private sector job | -0.111*** | -0.0733*** | 0.0187** | -0.0628 | -0.0918 |
|  | (0.0313) | (0.0229) | (0.00911) | (0.0694) | (0.0638) |
| Self employed | -0.117*** | -0.0788*** | 0.0178** | -0.0903 | -0.0763 |
|  | (0.0283) | (0.0207) | (0.00822) | (0.0627) | (0.0576) |
| Unemployed | -0.0749 | -0.138** | -0.0241 | -0.112 | -0.162 |
|  | (0.0909) | (0.0665) | (0.0264) | (0.202) | (0.185) |
| Economically inactive | -0.0589** | -0.0619*** | -0.000894 | -0.0574 | -0.0498 |
|  | (0.0293) | (0.0214) | (0.00852) | (0.0650) | (0.0597) |
| Constant | 7.320*** | 6.857*** | 0.658*** | -0.686*** | -1.014*** |
|  | (0.100) | (0.0731) | (0.0291) | (0.222) | (0.204) |
|  |  |  |  |  |  |
| Observations | 9,668 | 9,668 | 9,668 | 9,668 | 9,668 |
| R-squared | 0.436 | 0.497 | 0.257 | 0.277 | 0.386 |

Standard errors in parentheses
*** p<0.01, ** p<0.05, * p<0.1

| VARIABLES | (1) Ln(MPCE) | (2) Ln(MPCE food) | (3) Share of food expenditure | (4) Simpson | (5) Shannon |
|---|---|---|---|---|---|
| Table #: OLS estimates for households left behind (urban) | | | | | |
| household with no emigrant (base cat.) | | | | | |
| household with an emigrant | 0.246*** | 0.135*** | -0.0431*** | 0.136*** | 0.110*** |
| | (0.0208) | (0.0150) | (0.00601) | (0.0432) | (0.0423) |
| Dependency ratio | -0.000573*** | -0.000439*** | 3.46e-05 | 0.000376 | 0.000570* |
| | (0.000137) | (9.87e-05) | (3.96e-05) | (0.000285) | (0.000279) |
| Gender of household head (base cat: male) | | | | | |
| Female | 0.00188 | -0.00390 | 0.00134 | 0.0212 | -0.00148 |
| | (0.0311) | (0.0224) | (0.00899) | (0.0646) | (0.0633) |
| Age of household head (base cat: 15-24 years) | | | | | |
| 25-34 years | -0.0411 | 0.00894 | 0.0135 | 0.867*** | 0.962*** |
| | (0.158) | (0.114) | (0.0456) | (0.328) | (0.321) |
| 35-44 years | 0.0166 | -0.0230 | -0.0261 | 0.773** | 0.870*** |
| | (0.152) | (0.110) | (0.0441) | (0.317) | (0.310) |
| 45-59 years | 0.127 | 0.0317 | -0.0468 | 0.769** | 0.855*** |
| | (0.151) | (0.109) | (0.0438) | (0.315) | (0.308) |
| 60 and above years | 0.136 | 0.0228 | -0.0567 | 0.676** | 0.764** |
| | (0.151) | (0.109) | (0.0438) | (0.315) | (0.309) |
| Religion of household head (base cat: Hindu) | | | | | |
| Christian | 0.0585*** | 0.0783*** | 0.00980 | 0.112** | 0.132*** |
| | (0.0211) | (0.0152) | (0.00611) | (0.0439) | (0.0430) |
| Muslim | -0.0281 | 0.0120 | 0.0143** | -0.000143 | 0.0324 |
| | (0.0226) | (0.0163) | (0.00653) | (0.0469) | (0.0460) |
| Education level of household head (base cat: Primary) | | | | | |
| Upper Primary | 0.0226 | 0.00724 | -0.00797 | 0.147*** | 0.109** |
| | (0.0259) | (0.0187) | (0.00749) | (0.0538) | (0.0527) |
| Secondary | 0.173*** | 0.0765*** | -0.0422*** | 0.126** | 0.115** |
| | (0.0240) | (0.0173) | (0.00695) | (0.0499) | (0.0489) |
| Higher Secondary | 0.383*** | 0.153*** | -0.100*** | 0.329*** | 0.368*** |
| | (0.0425) | (0.0306) | (0.0123) | (0.0883) | (0.0865) |
| Degree and above | 0.461*** | 0.220*** | -0.0966*** | 0.279*** | 0.317*** |
| | (0.0312) | (0.0225) | (0.00901) | (0.0648) | (0.0635) |
| not educated | -0.127*** | -0.126*** | -0.000782 | -0.120 | -0.139* |
| | (0.0357) | (0.0257) | (0.0103) | (0.0742) | (0.0727) |
| Literate without school education | -0.0337 | -0.0826* | -0.0147 | 0.419*** | 0.433*** |
| | (0.0650) | (0.0469) | (0.0188) | (0.135) | (0.133) |
| Marital status of household head (base cat: Unmarried) | | | | | |
| Married | 0.0208 | 0.0700 | 0.00916 | 0.244* | 0.266* |
| | (0.0685) | (0.0494) | (0.0198) | (0.142) | (0.140) |
| Divorced and others | 0.00751 | 0.0665 | 0.0161 | 0.385*** | 0.390*** |
| | (0.0703) | (0.0507) | (0.0203) | (0.146) | (0.143) |
| Size of household (base cat: 0-2) | | | | | |
| 3 - 5 members | -0.318*** | -0.308*** | -0.00132 | 0.155*** | 0.276*** |
| | (0.0224) | (0.0161) | (0.00648) | (0.0466) | (0.0456) |
| 6-10 members | -0.554*** | -0.511*** | 0.0149* | -0.201*** | 0.0987* |
| | (0.0281) | (0.0202) | (0.00812) | (0.0584) | (0.0572) |
| 10 and above members | -0.806*** | -0.690*** | 0.0482** | -0.796*** | -0.115 |
| | (0.0838) | (0.0604) | (0.0243) | (0.174) | (0.171) |
| Land holdings in acres (base cat: no land) | | | | | |
| 0.01-0.100 | 0.0808** | 0.0601*** | -0.00657 | 0.171** | 0.142** |
| | (0.0319) | (0.0230) | (0.00924) | (0.0664) | (0.0650) |
| 0.101-0..200 | 0.186*** | 0.145*** | -0.0122 | 0.177** | 0.186** |
|                                                    | (0.0431)  | (0.0311)  | (0.0125)  | (0.0897)  | (0.0879)  |
| 0.201-0.300                                        | 0.0789    | 0.0916**  | -0.00217  | 0.0927    | 0.142     |
|                                                    | (0.0521)  | (0.0376)  | (0.0151)  | (0.108)   | (0.106)   |
| 0.301-.0.500                                       | 0.0984*   | 0.0965**  | -0.0112   | -0.0857   | -0.0696   |
|                                                    | (0.0520)  | (0.0375)  | (0.0151)  | (0.108)   | (0.106)   |
| 0.501-1.000                                        | 0.0852*   | 0.0708**  | -0.0150   | 0.00841   | -0.0652   |
|                                                    | (0.0501)  | (0.0361)  | (0.0145)  | (0.104)   | (0.102)   |
| 1.001-3.000                                        | 0.364***  | 0.238***  | -0.0693***| 0.225     | 0.275*    |
|                                                    | (0.0718)  | (0.0518)  | (0.0208)  | (0.149)   | (0.146)   |
| More than 3.000                                    | 0.287**   | 0.134     | -0.0744** | 0.259     | 0.249     |
|                                                    | (0.121)   | (0.0871)  | (0.0350)  | (0.251)   | (0.246)   |
| Economic activity of household head (base cat: Govt. job) |    |           |           |           |           |
| Private sector job                                 | -0.0391   | -0.00593  | 0.0198    | -0.0209   | 0.00749   |
|                                                    | (0.0449)  | (0.0324)  | (0.0130)  | (0.0934)  | (0.0915)  |
| Self employed                                      | -0.103**  | -0.0359   | 0.0352*** | 0.0862    | 0.0742    |
|                                                    | (0.0414)  | (0.0299)  | (0.0120)  | (0.0862)  | (0.0845)  |
| Unemployed                                         | -0.0249   | -0.00838  | 0.00571   | 0.296     | 0.0890    |
|                                                    | (0.156)   | (0.113)   | (0.0452)  | (0.325)   | (0.318)   |
| Economically inactive                              | -0.0812*  | -0.0153   | 0.0361*** | 0.0771    | 0.0675    |
|                                                    | (0.0434)  | (0.0313)  | (0.0125)  | (0.0902)  | (0.0884)  |
| Constant                                           | 7.603***  | 6.960***  | 0.560***  | -1.335*** | -1.558*** |
|                                                    | (0.168)   | (0.121)   | (0.0485)  | (0.349)   | (0.342)   |
|                                                    |           |           |           |           |           |
| Observations                                       | 2,762     | 2,762     | 2,762     | 2,762     | 2,762     |
| R-squared                                          | 0.473     | 0.529     | 0.247     | 0.351     | 0.396     |
| Standard errors in parentheses                     |           |           |           |           |           |
| *** p<0.01, ** p<0.05, * p<0.1                     |           |           |           |           |           |

| | (1) | (2) | (3) | (4) | (5) |
|---|---|---|---|---|---|
| VARIABLES | Ln(MPCE) | Ln(MPCE food) | Share of food expenditure | Simpson | Shannon |
| household with no emigrant (base cat.) | | | | | |
| household with an emigrant | 0.594*** | 0.295*** | -0.130*** | 0.625*** | 0.755*** |
| | (0.0697) | (0.0577) | (0.0179) | (0.174) | (0.174) |
| Dependency ratio | -0.000971*** | -0.000645*** | 0.000129*** | -0.000158 | -0.000154 |
| | (8.68e-05) | (6.49e-05) | (2.29e-05) | (0.000214) | (0.000203) |
| Gender of household head (base cat: male) | | | | | |
| Female | -0.114*** | -0.0294 | 0.0374*** | -0.164** | -0.183** |
| | (0.0298) | (0.0240) | (0.00743) | (0.0706) | (0.0755) |
| Age of household head (base cat: 15-24 years) | | | | | |
| 25-34 years | -0.0419 | -0.0273 | 0.0128 | -0.189 | -0.186 |
| | (0.0856) | (0.0697) | (0.0252) | (0.183) | (0.182) |
| 35-44 years | -0.0645** | -0.00104 | 0.0288*** | 0.152*** | 0.160*** |
| | (0.0256) | (0.0194) | (0.00684) | (0.0589) | (0.0585) |
| 45-59 years | -0.0403*** | -0.0145 | 0.0112*** | 0.0359 | 0.0439 |
| | (0.0153) | (0.0125) | (0.00422) | (0.0374) | (0.0381) |
| 60 and above years | 0.0110 | 0.00431 | -0.00231 | 0.0178 | 0.0116 |
| | (0.0108) | (0.00796) | (0.00327) | (0.0233) | (0.0234) |
| Religion of household head (base cat: Hindu) | | | | | |
| Christian | 0.0701*** | 0.0698*** | 0.00108 | -0.0498 | -0.0465 |
| | (0.0190) | (0.0152) | (0.00493) | (0.0455) | (0.0451) |
| Muslim | -0.0741*** | 0.0106 | 0.0358*** | -0.116 | -0.110 |
| | (0.0273) | (0.0217) | (0.00664) | (0.0731) | (0.0705) |
| Education level of household head (base cat: Primary) | | | | | |
| Upper Primary | 0.0390*** | 0.0189** | -0.0104*** | 0.0589** | 0.0737*** |
| | (0.0130) | (0.00904) | (0.00373) | (0.0277) | (0.0264) |
| Secondary | 0.145*** | 0.0808*** | -0.0305*** | 0.0651** | 0.0867*** |
| | (0.0153) | (0.0113) | (0.00438) | (0.0330) | (0.0332) |
| Higher Secondary | 0.315*** | 0.157*** | -0.0661*** | 0.136** | 0.172*** |
| | (0.0282) | (0.0189) | (0.00778) | (0.0575) | (0.0553) |
| Degree and above | 0.430*** | 0.247*** | -0.0801*** | 0.215*** | 0.299*** |
| | (0.0244) | (0.0176) | (0.00707) | (0.0588) | (0.0636) |
| not educated | -0.0818*** | -0.0870*** | -0.00256 | -0.0647 | -0.0493 |
| | (0.0193) | (0.0147) | (0.00570) | (0.0413) | (0.0383) |
| Literate without school education | -0.0456 | -0.0203 | 0.0117 | 0.134 | 0.149 |
| | (0.0425) | (0.0283) | (0.00999) | (0.124) | (0.119) |
| Marital status of household head (base cat: Unmarried) | | | | | |
| Married | -0.0995** | -0.0205 | 0.0305** | 0.0245 | -0.0273 |
| | (0.0435) | (0.0340) | (0.0127) | (0.0995) | (0.0946) |
| Divorced and others | -0.00676 | 0.0201 | 0.00773 | 0.229*** | 0.208** |
| | (0.0385) | (0.0278) | (0.0117) | (0.0851) | (0.0846) |
| Size of household (base cat: 0-2) | | | | | |
| 3 - 5 members | -0.246*** | -0.285*** | -0.0247*** | 0.0549* | 0.188*** |
| | (0.0147) | (0.0123) | (0.00422) | (0.0320) | (0.0332) |
| 6-10 members | -0.457*** | -0.472*** | -0.0146*** | -0.319*** | -0.0207 |
| | (0.0199) | (0.0167) | (0.00534) | (0.0489) | (0.0460) |
| 10 and above members | -0.734*** | -0.678*** | 0.0143 | -0.526*** | -0.112 |
| | (0.0470) | (0.0331) | (0.0123) | (0.139) | (0.115) |
| Land holdings in acres (base cat: no land) | | | | | |
| 0.01-0.100 | 0.0925*** | 0.0491** | -0.0223*** | 0.0769 | 0.0771 |

Table A4: Instrumental Variable estimates for households left behind (rural + urban)

|  | (0.0316) | (0.0220) | (0.00713) | (0.0974) | (0.0845) |
| --- | --- | --- | --- | --- | --- |
| 0.101-0..200 | 0.0698** | 0.0590** | -0.00660 | 0.0731 | 0.0611 |
|  | (0.0352) | (0.0238) | (0.00816) | (0.0803) | (0.0675) |
| 0.201-0.300 | 0.0339 | 0.0456** | 0.00371 | -0.00116 | -0.0167 |
|  | (0.0292) | (0.0211) | (0.00859) | (0.0686) | (0.0649) |
| 0.301-.0.500 | 0.0403 | 0.0630*** | 0.00523 | 0.0121 | -0.0232 |
|  | (0.0295) | (0.0211) | (0.00720) | (0.0735) | (0.0654) |
| 0.501-1.000 | 0.152*** | 0.109*** | -0.0224*** | 0.124* | 0.0846 |
|  | (0.0312) | (0.0223) | (0.00843) | (0.0663) | (0.0605) |
| 1.001-3.000 | 0.183*** | 0.111*** | -0.0327*** | 0.0922 | 0.0737 |
|  | (0.0341) | (0.0211) | (0.00903) | (0.0799) | (0.0733) |
| More than 3.000 | 0.208*** | 0.138*** | -0.0312** | 0.171* | 0.129 |
|  | (0.0463) | (0.0256) | (0.0128) | (0.0947) | (0.0882) |
| Economic activity of household head (base cat: Govt. job) | | | | | |
| Private sector job | -0.0820*** | -0.0518** | 0.0161* | -0.0270 | -0.0244 |
|  | (0.0303) | (0.0227) | (0.00826) | (0.0643) | (0.0671) |
| Self employed | -0.126*** | -0.0766*** | 0.0247*** | -0.0687 | -0.0595 |
|  | (0.0250) | (0.0187) | (0.00709) | (0.0546) | (0.0509) |
| Unemployed | -0.166** | -0.153** | 0.00976 | -0.0949 | -0.200 |
|  | (0.0827) | (0.0607) | (0.0219) | (0.184) | (0.167) |
| Economically inactive | -0.117*** | -0.0736*** | 0.0217*** | -0.0972 | -0.108* |
|  | (0.0280) | (0.0220) | (0.00823) | (0.0658) | (0.0632) |
| Constant | 7.570*** | 6.947*** | 0.572*** | -0.364** | -0.656*** |
|  | (0.0715) | (0.0536) | (0.0203) | (0.163) | (0.189) |
|  |  |  |  |  |  |
| Observations | 12,431 | 12,431 | 12,431 | 12,431 | 12,431 |
| R-squared | 0.373 | 0.469 | 0.178 | 0.247 | 0.316 |

Robust standard errors in parentheses
*** p<0.01, ** p<0.05, * p<0.1

Table A5: Instrumental Variable estimates for households left behind (rural)

| VARIABLES | (1) Ln(MPCE) | (2) Ln(MPCE food) | (3) Share of food expenditure | (4) Simpson | (5) Shannon |
|---|---|---|---|---|---|
| household with no emigrant (base cat.) | | | | | |
| household with an emigrant | 0.582*** | 0.284*** | -0.133*** | 0.472** | 0.570*** |
|  | (0.0772) | (0.0643) | (0.0196) | (0.187) | (0.179) |
| Dependency ratio | -0.000979*** | -0.000645*** | 0.000136*** | -6.31e-06 | 1.85e-06 |
|  | (9.46e-05) | (7.26e-05) | (2.44e-05) | (0.000235) | (0.000217) |
| Gender of household head (base cat: male) | | | | | |
| Female | -0.105*** | -0.0142 | 0.0418*** | -0.101 | -0.107 |
|  | (0.0329) | (0.0272) | (0.00785) | (0.0719) | (0.0683) |
| Age of household head (base cat: 15-24 years) | | | | | |
| 25-34 years | -0.0381 | -0.0316 | 0.0115 | -0.0707 | -0.00815 |
|  | (0.100) | (0.0859) | (0.0299) | (0.190) | (0.168) |
| 35-44 years | -0.0289 | 0.00840 | 0.0173** | 0.150** | 0.154** |
|  | (0.0283) | (0.0217) | (0.00756) | (0.0621) | (0.0601) |
| 45-59 years | -0.0131 | -0.00233 | 0.00459 | 0.0194 | 0.0263 |
|  | (0.0179) | (0.0140) | (0.00494) | (0.0390) | (0.0371) |
| 60 and above years | 0.0242* | 0.00765 | -0.00726* | 0.0113 | 0.00623 |
|  | (0.0124) | (0.00934) | (0.00371) | (0.0242) | (0.0229) |
| Religion of household head (base cat: Hindu) | | | | | |
| Christian | 0.0898*** | 0.0785*** | -0.00409 | -0.0629 | -0.0635 |
|  | (0.0215) | (0.0164) | (0.00574) | (0.0504) | (0.0472) |
| Muslim | -0.0556* | 0.0312 | 0.0378*** | -0.0894 | -0.0826 |
|  | (0.0305) | (0.0246) | (0.00708) | (0.0793) | (0.0690) |
| Education level of household head (base cat: Primary) | | | | | |
| Upper Primary | 0.0419*** | 0.0212** | -0.0108*** | 0.0395 | 0.0582** |
|  | (0.0143) | (0.0101) | (0.00413) | (0.0292) | (0.0268) |
| Secondary | 0.130*** | 0.0747*** | -0.0270*** | 0.0458 | 0.0681** |
|  | (0.0172) | (0.0125) | (0.00493) | (0.0352) | (0.0334) |
| Higher Secondary | 0.292*** | 0.151*** | -0.0585*** | 0.0925 | 0.120** |
|  | (0.0325) | (0.0215) | (0.00894) | (0.0631) | (0.0568) |
| Degree and above | 0.387*** | 0.235*** | -0.0707*** | 0.142** | 0.214*** |
|  | (0.0287) | (0.0195) | (0.00854) | (0.0570) | (0.0564) |
| not educated | -0.0735*** | -0.0786*** | -0.00246 | -0.0672 | -0.0490 |
|  | (0.0218) | (0.0165) | (0.00640) | (0.0446) | (0.0398) |
| Literate without school education | -0.0600 | -0.0227 | 0.0152 | 0.0484 | 0.0707 |
|  | (0.0458) | (0.0286) | (0.0113) | (0.110) | (0.105) |
| Marital status of household head (base cat: Unmarried) | | | | | |
| Married | -0.106** | -0.0338 | 0.0304** | 0.0610 | 0.00959 |
|  | (0.0496) | (0.0387) | (0.0145) | (0.110) | (0.0994) |
| Divorced and others | -0.0220 | -0.00467 | 0.00509 | 0.176* | 0.143 |
|  | (0.0433) | (0.0312) | (0.0137) | (0.0962) | (0.0875) |
| Size of household (base cat: 0-2) | | | | | |
| 3 - 5 members | -0.236*** | -0.284*** | -0.0290*** | 0.00440 | 0.132*** |
|  | (0.0168) | (0.0143) | (0.00468) | (0.0350) | (0.0353) |
| 6-10 members | -0.447*** | -0.471*** | -0.0189*** | -0.387*** | -0.0993* |
|  | (0.0224) | (0.0191) | (0.00588) | (0.0550) | (0.0514) |
| 10 and above members | -0.721*** | -0.685*** | 0.00311 | -0.471*** | -0.0825 |
|  | (0.0518) | (0.0364) | (0.0140) | (0.145) | (0.122) |
| Land holdings in acres (base cat: no land) | | | | | |
| 0.01-0.100 | 0.0916** | 0.0427* | -0.0265*** | 0.0609 | 0.0689 |
| | (0.0360) | (0.0252) | (0.00828) | (0.117) | (0.100) |
| 0.101-0..200 | 0.0505 | 0.0394 | -0.00877 | 0.0730 | 0.0675 |
| | (0.0400) | (0.0260) | (0.00953) | (0.0930) | (0.0758) |
| 0.201-0.300 | 0.0302 | 0.0358 | 0.00206 | -0.00562 | -0.0234 |
| | (0.0340) | (0.0246) | (0.0101) | (0.0780) | (0.0715) |
| 0.301-.0.500 | 0.0468 | 0.0661*** | 0.00443 | 0.0566 | 0.0254 |
| | (0.0320) | (0.0225) | (0.00804) | (0.0829) | (0.0715) |
| 0.501-1.000 | 0.163*** | 0.117*** | -0.0233** | 0.162** | 0.137** |
| | (0.0326) | (0.0233) | (0.00962) | (0.0747) | (0.0661) |
| 1.001-3.000 | 0.180*** | 0.106*** | -0.0322*** | 0.123 | 0.109 |
| | (0.0359) | (0.0222) | (0.00967) | (0.0860) | (0.0764) |
| More than 3.000 | 0.218*** | 0.150*** | -0.0295** | 0.216** | 0.189** |
| | (0.0491) | (0.0278) | (0.0136) | (0.0975) | (0.0896) |
| Economic activity of household head (base cat: Govt. job) | | | | | |
| Private sector job | -0.118*** | -0.0759*** | 0.0204** | -0.0688 | -0.0993 |
| | (0.0348) | (0.0261) | (0.00980) | (0.0694) | (0.0649) |
| Self employed | -0.132*** | -0.0849*** | 0.0216** | -0.104 | -0.0933 |
| | (0.0300) | (0.0221) | (0.00880) | (0.0649) | (0.0584) |
| Unemployed | -0.186* | -0.184** | 0.00467 | -0.215 | -0.291 |
| | (0.105) | (0.0739) | (0.0254) | (0.222) | (0.196) |
| Economically inactive | -0.122*** | -0.0884*** | 0.0155 | -0.117 | -0.124* |
| | (0.0334) | (0.0255) | (0.00993) | (0.0766) | (0.0707) |
| Constant | 7.416*** | 6.883*** | 0.617*** | -0.558*** | -0.953*** |
| | (0.0758) | (0.0617) | (0.0232) | (0.170) | (0.191) |
| | | | | | |
| Observations | 9,668 | 9,668 | 9,668 | 9,668 | 9,668 |
| R-squared | 0.381 | 0.480 | 0.199 | 0.264 | 0.366 |

Robust standard errors in parentheses
*** p<0.01, ** p<0.05, * p<0.1

| | (1) | (2) | (3) | (4) | (5) |
|---|---|---|---|---|---|
| VARIABLES | Ln(MPCE) | Ln(MPCE food) | Share of food expenditure | Simpson | Shannon |
| Table A6: Instrumental Variable estimates for households left behind (urban) | | | | | |
| household with no emigrant (base cat.) | | | | | |
| household with an emigrant | 0.567*** | 0.244** | -0.131*** | 0.941** | 1.095** |
|  | (0.137) | (0.115) | (0.0349) | (0.374) | (0.454) |
| Dependency ratio | -0.000831*** | -0.000526*** | 0.000105** | -0.000271 | -0.000221 |
|  | (0.000174) | (0.000130) | (5.19e-05) | (0.000403) | (0.000440) |
| Gender of household head (base cat: male) | | | | | |
| Female | -0.104* | -0.0308 | 0.0265* | -0.281* | -0.316 |
|  | (0.0602) | (0.0428) | (0.0151) | (0.162) | (0.202) |
| Age of household head (base cat: 15-24 years) | | | | | |
| 25-34 years | -0.0506 | 0.00484 | 0.0318 | -0.525 | -0.575 |
|  | (0.156) | (0.123) | (0.0480) | (0.467) | (0.463) |
| 35-44 years | -0.133** | 0.000937 | 0.0577*** | 0.290** | 0.320** |
|  | (0.0553) | (0.0408) | (0.0138) | (0.135) | (0.142) |
| 45-59 years | -0.0853*** | -0.0344 | 0.0211*** | 0.177** | 0.203** |
|  | (0.0285) | (0.0219) | (0.00787) | (0.0853) | (0.0978) |
| 60 and above years | -0.000928 | 0.0115 | 0.00762 | 0.109** | 0.111** |
|  | (0.0234) | (0.0151) | (0.00672) | (0.0507) | (0.0539) |
| Religion of household head (base cat: Hindu) | | | | | |
| Christian | 0.0358 | 0.0706** | 0.0160* | 0.0545 | 0.0619 |
|  | (0.0326) | (0.0278) | (0.00864) | (0.0852) | (0.0978) |
| Muslim | -0.104** | -0.0136 | 0.0350** | -0.189 | -0.198 |
|  | (0.0528) | (0.0325) | (0.0140) | (0.136) | (0.162) |
| Education level of household head (base cat: Primary) | | | | | |
| Upper Primary | 0.0234 | 0.00754 | -0.00816 | 0.150** | 0.113** |
|  | (0.0281) | (0.0164) | (0.00806) | (0.0589) | (0.0540) |
| Secondary | 0.154*** | 0.0700*** | -0.0369*** | 0.0796 | 0.0580 |
|  | (0.0230) | (0.0175) | (0.00734) | (0.0692) | (0.0725) |
| Higher Secondary | 0.347*** | 0.141*** | -0.0903*** | 0.243** | 0.261** |
|  | (0.0442) | (0.0346) | (0.0126) | (0.0989) | (0.103) |
| Degree and above | 0.429*** | 0.209*** | -0.0876*** | 0.206* | 0.227* |
|  | (0.0381) | (0.0286) | (0.0115) | (0.120) | (0.132) |
| not educated | -0.112*** | -0.121*** | -0.00493 | -0.0824 | -0.0928 |
|  | (0.0307) | (0.0265) | (0.0101) | (0.104) | (0.116) |
| Literate without school education | -0.0323 | -0.0823** | -0.0152 | 0.414* | 0.427* |
|  | (0.0412) | (0.0410) | (0.0140) | (0.219) | (0.221) |
| Marital status of household head (base cat: Unmarried) | | | | | |
| Married | -0.0563 | 0.0428 | 0.0289 | -0.00277 | -0.0321 |
|  | (0.0989) | (0.0759) | (0.0236) | (0.216) | (0.235) |
| Divorced and others | 0.0405 | 0.0767 | 0.00594 | 0.419** | 0.435** |
|  | (0.0895) | (0.0640) | (0.0204) | (0.170) | (0.190) |
| Size of household (base cat: 0-2) | | | | | |
| 3 - 5 members | -0.291*** | -0.298*** | -0.00875 | 0.228*** | 0.366*** |
|  | (0.0252) | (0.0210) | (0.00797) | (0.0711) | (0.0749) |
| 6-10 members | -0.512*** | -0.497*** | 0.00342 | -0.0924 | 0.231** |
|  | (0.0380) | (0.0284) | (0.0109) | (0.0980) | (0.0942) |
| 10 and above members | -0.809*** | -0.690*** | 0.0489* | -0.801** | -0.121 |
|  | (0.0871) | (0.0561) | (0.0263) | (0.404) | (0.309) |
| Land holdings in acres (base cat: no land) | | | | | |
| 0.01-0.100 | 0.0661 | 0.0551* | -0.00259 | 0.132 | 0.0956 |

|  | (0.0482) | (0.0312) | (0.0108) | (0.0852) | (0.0910) |
|---|---|---|---|---|---|
| 0.101-0..200 | 0.134** | 0.127*** | 0.00180 | 0.0423 | 0.0215 |
|  | (0.0640) | (0.0447) | (0.0140) | (0.135) | (0.142) |
| 0.201-0.300 | 0.0625 | 0.0860*** | 0.00226 | 0.0492 | 0.0889 |
|  | (0.0464) | (0.0332) | (0.0129) | (0.114) | (0.116) |
| 0.301-.0.500 | 0.0677 | 0.0860 | -0.00286 | -0.164 | -0.165 |
|  | (0.0709) | (0.0556) | (0.0124) | (0.114) | (0.129) |
| 0.501-1.000 | 0.0770 | 0.0680 | -0.0130 | -0.0179 | -0.0970 |
|  | (0.0961) | (0.0661) | (0.0200) | (0.0993) | (0.121) |
| 1.001-3.000 | 0.366*** | 0.239*** | -0.0699*** | 0.229 | 0.279* |
|  | (0.0843) | (0.0455) | (0.0225) | (0.140) | (0.166) |
| More than 3.000 | 0.190** | 0.101 | -0.0482** | 0.0136 | -0.0506 |
|  | (0.0928) | (0.0665) | (0.0215) | (0.245) | (0.270) |
| Economic activity of household head (base cat: Govt. job) | | | | | |
| Private sector job | -0.0431 | -0.00726 | 0.0210 | -0.0292 | -0.00283 |
|  | (0.0476) | (0.0337) | (0.0139) | (0.116) | (0.117) |
| Self employed | -0.117** | -0.0408 | 0.0392*** | 0.0532 | 0.0336 |
|  | (0.0459) | (0.0323) | (0.0124) | (0.102) | (0.0909) |
| Unemployed | -0.153 | -0.0451 | 0.0484 | 0.294 | 0.0642 |
|  | (0.133) | (0.0938) | (0.0469) | (0.261) | (0.270) |
| Economically inactive | -0.134*** | -0.0332 | 0.0505*** | -0.0551 | -0.0942 |
|  | (0.0475) | (0.0372) | (0.0143) | (0.122) | (0.121) |
| Constant | 7.794*** | 6.998*** | 0.491*** | -0.453 | -0.573* |
|  | (0.119) | (0.0915) | (0.0280) | (0.286) | (0.320) |
| | | | | | |
| Observations | 2,763 | 2,763 | 2,763 | 2,763 | 2,763 |
| R-squared | 0.427 | 0.520 | 0.188 | 0.268 | 0.275 |

Robust standard errors in parentheses
*** p<0.01, ** p<0.05, * p<0.1

Table A4: Sub-group wise expenditure OLS estimates for left-behind households (rural+ urban)

| Variables | (1) Cereals | (2) Pulses | (3) Milk and related products | (4) Oil, ghee and butter etc. | (5) Vegetables and fruits | (6) Egg, fish and meat | (7) Sugar | (8) Processed food | (9) Ready to eat | (10) Tobacco, liquor and intoxicants |
|---|---|---|---|---|---|---|---|---|---|---|
| | | | | MPCE spendings on various sub-groups | | | | | | |
| household with no emigrant (base cat.) | | | | | | | | | | |
| household with an emigrant | 0.119*** | 0.0606*** | 0.201*** | 0.110*** | 0.173*** | 0.218*** | 0.0674*** | 0.318*** | -0.0157 | -0.224*** |
| | (0.0103) | (0.0101) | (0.0146) | (0.0107) | (0.0127) | (0.0239) | (0.0105) | (0.0312) | (0.0382) | (0.0488) |
| Dependency ratio | -0.000492*** | -0.000688*** | -0.000228** | -0.000546*** | -0.000430*** | -0.000496*** | -0.000625*** | 0.000533*** | -0.00101*** | -0.00122*** |
| | (6.46e-05) | (6.30e-05) | (9.10e-05) | (6.66e-05) | (7.93e-05) | (0.000149) | (6.54e-05) | (0.000195) | (0.000238) | (0.000305) |
| Gender of household head (base cat: male) | | | | | | | | | | |
| Female | 0.0266* | 0.0446*** | 0.00140 | 0.0274* | 0.0606*** | 0.0376 | 0.0224 | 0.0521 | -0.242*** | -0.540*** |
| | (0.0151) | (0.0148) | (0.0213) | (0.0156) | (0.0186) | (0.0350) | (0.0153) | (0.0456) | (0.0558) | (0.0714) |
| Age of household head (base cat: 15-24 years) | | | | | | | | | | |
| 25-34 years | -0.0908 | -0.0317 | 0.171 | 0.0991 | 0.151 | 0.180 | -0.0699 | -0.0933 | 0.0881 | 0.0474 |
| | (0.0837) | (0.0817) | (0.118) | (0.0864) | (0.103) | (0.194) | (0.0849) | (0.253) | (0.309) | (0.395) |
| 35-44 years | -0.0597 | -0.0356 | 0.0436 | 0.0944 | 0.150 | 0.165 | -0.0638 | -0.166 | -0.0710 | -0.0504 |
| | (0.0821) | (0.0802) | (0.116) | (0.0847) | (0.101) | (0.190) | (0.0833) | (0.248) | (0.303) | (0.388) |
| 45-59 years | -0.0338 | -0.0135 | 0.109 | 0.162* | 0.180* | 0.172 | -0.0588 | -0.166 | -0.0264 | -0.0677 |
| | (0.0818) | (0.0799) | (0.115) | (0.0844) | (0.101) | (0.189) | (0.0829) | (0.247) | (0.302) | (0.386) |
| 60 and above years | -0.0276 | -0.00148 | 0.114 | 0.155* | 0.198** | 0.0749 | -0.0520 | -0.256 | -0.176 | -0.184 |
| | (0.0820) | (0.0800) | (0.116) | (0.0846) | (0.101) | (0.189) | (0.0831) | (0.247) | (0.303) | (0.387) |
| Religion of household head (base cat: Hindu) | | | | | | | | | | |
| Christian | 0.0666*** | 0.0303*** | 0.0761*** | 0.0607*** | -0.0163 | 0.490*** | 0.0450*** | 0.0114 | 0.0897** | -0.244*** |
| | (0.0102) | (0.00992) | (0.0143) | (0.0105) | (0.0125) | (0.0235) | (0.0103) | (0.0307) | (0.0375) | (0.0480) |
| Muslim | 0.0819*** | -0.0128 | -0.00185 | -0.00488 | -0.0462*** | 0.401*** | 0.0309*** | 0.0256 | -0.0162 | -0.911*** |
| | (0.0109) | (0.0106) | (0.0153) | (0.0112) | (0.0134) | (0.0252) | (0.0110) | (0.0328) | (0.0402) | (0.0514) |
| Education level of household head (base cat: Primary) | | | | | | | | | | |
| Upper Primary | 0.0417*** | 0.0536*** | 0.0817*** | 0.00884 | 0.0268* | -0.0124 | 0.00975 | 0.134*** | -0.0348 | -0.232*** |
| | (0.0114) | (0.0112) | (0.0161) | (0.0118) | (0.0140) | (0.0264) | (0.0116) | (0.0345) | (0.0422) | (0.0540) |
| Secondary | 0.121*** | 0.111*** | 0.166*** | 0.0798*** | 0.119*** | -0.0573** | 0.0480*** | 0.264*** | -0.0372 | -0.475*** |
|  | (0.0112) | (0.0109) | (0.0157) | (0.0115) | (0.0137) | (0.0258) | (0.0113) | (0.0336) | (0.0412) | (0.0526) |
| Higher Secondary | 0.185*** | 0.155*** | 0.324*** | 0.144*** | 0.243*** | -0.116** | 0.0829*** | 0.465*** | 0.0656 | -0.849*** |
|  | (0.0203) | (0.0198) | (0.0286) | (0.0209) | (0.0249) | (0.0469) | (0.0206) | (0.0613) | (0.0750) | (0.0958) |
| Degree and above | 0.252*** | 0.214*** | 0.501*** | 0.168*** | 0.330*** | -0.235*** | 0.162*** | 0.677*** | 0.288*** | -1.117*** |
|  | (0.0177) | (0.0172) | (0.0249) | (0.0182) | (0.0217) | (0.0408) | (0.0179) | (0.0533) | (0.0652) | (0.0833) |
| not educated | -0.104*** | -0.0566*** | -0.104*** | -0.0801*** | -0.143*** | -0.0396 | -0.111*** | -0.106** | 0.0104 | 0.472*** |
|  | (0.0151) | (0.0147) | (0.0212) | (0.0155) | (0.0185) | (0.0348) | (0.0153) | (0.0455) | (0.0556) | (0.0711) |
| Literate without school education | -0.0282 | 0.0251 | 0.00169 | -0.0853*** | -0.0235 | -0.0133 | 0.0474 | -0.0320 | 0.279*** | 0.254* |
|  | (0.0287) | (0.0280) | (0.0404) | (0.0296) | (0.0352) | (0.0662) | (0.0290) | (0.0864) | (0.106) | (0.135) |
| Marital status of household head (base cat: Unmarried) |  |  |  |  |  |  |  |  |  |  |
| Married | 0.109*** | -0.0745** | -0.125** | 0.0523 | 0.266*** | 0.236*** | -0.0508 | 0.437*** | -0.504*** | -0.189 |
|  | (0.0353) | (0.0344) | (0.0497) | (0.0364) | (0.0433) | (0.0815) | (0.0357) | (0.106) | (0.130) | (0.166) |
| Divorced and others | 0.0571 | -0.0728** | -0.111** | 0.0283 | 0.210*** | 0.164** | -0.0344 | 0.439*** | -0.328** | -0.265 |
|  | (0.0360) | (0.0351) | (0.0507) | (0.0371) | (0.0442) | (0.0831) | (0.0365) | (0.109) | (0.133) | (0.170) |
| Size of household (base cat: 0-2) |  |  |  |  |  |  |  |  |  |  |
| 3 - 5 members | -0.166*** | -0.542*** | -0.437*** | -0.328*** | -0.252*** | 0.00215 | -0.301*** | 0.376*** | 0.329*** | 0.294*** |
|  | (0.0110) | (0.0107) | (0.0154) | (0.0113) | (0.0135) | (0.0253) | (0.0111) | (0.0331) | (0.0405) | (0.0517) |
| 6-10 members | -0.114*** | -0.885*** | -0.716*** | -0.570*** | -0.522*** | -0.234*** | -0.524*** | 0.356*** | 0.421*** | 0.611*** |
|  | (0.0136) | (0.0133) | (0.0192) | (0.0140) | (0.0167) | (0.0314) | (0.0138) | (0.0410) | (0.0502) | (0.0642) |
| 10 and above members | -0.209*** | -1.159*** | -1.021*** | -0.794*** | -0.764*** | -0.353*** | -0.662*** | 0.330*** | 0.489*** | 1.045*** |
|  | (0.0366) | (0.0357) | (0.0516) | (0.0377) | (0.0449) | (0.0846) | (0.0371) | (0.110) | (0.135) | (0.173) |
| Land holdings in acres (base cat: no land) |  |  |  |  |  |  |  |  |  |  |
| 0.01-0.100 | -0.0153 | 0.0464*** | 0.101*** | 0.0390** | 0.110*** | 0.0202 | 0.0427*** | 0.162*** | 0.198*** | 0.0833 |
|  | (0.0157) | (0.0153) | (0.0222) | (0.0162) | (0.0193) | (0.0363) | (0.0159) | (0.0474) | (0.0580) | (0.0742) |
| 0.101-0..200 | 0.0406** | 0.0676*** | 0.0802*** | 0.0813*** | 0.0926*** | 0.0332 | 0.0761*** | 0.0802 | 0.159** | 0.115 |
|  | (0.0202) | (0.0197) | (0.0284) | (0.0208) | (0.0248) | (0.0466) | (0.0205) | (0.0609) | (0.0745) | (0.0952) |
| 0.201-0.300 | 0.0386* | 0.0325 | 0.109*** | 0.0998*** | 0.111*** | -0.0529 | 0.0580** | -0.0232 | -0.0121 | 0.0153 |
|  | (0.0226) | (0.0220) | (0.0318) | (0.0233) | (0.0277) | (0.0522) | (0.0229) | (0.0681) | (0.0834) | (0.107) |
| 0.301-.0.500 | 0.0750*** | 0.0461** | 0.149*** | 0.0808*** | 0.130*** | -0.0423 | 0.0364* | 0.0165 | -0.0170 | -0.198** |
|  | (0.0195) | (0.0190) | (0.0275) | (0.0201) | (0.0240) | (0.0451) | (0.0198) | (0.0588) | (0.0720) | (0.0920) |

| | (1) | (2) | (3) | (4) | (5) | (6) | (7) | (8) | (9) | (10) |
|---|---|---|---|---|---|---|---|---|---|---|
| 0.501-1.000 | 0.101*** | 0.0683*** | 0.235*** | 0.142*** | 0.206*** | -0.0119 | 0.102*** | 0.0831 | -0.0697 | -0.326*** |
| | (0.0187) | (0.0183) | (0.0264) | (0.0193) | (0.0230) | (0.0432) | (0.0190) | (0.0564) | (0.0691) | (0.0883) |
| 1.001-3.000 | 0.122*** | 0.104*** | 0.227*** | 0.109*** | 0.151*** | -0.0286 | 0.0936*** | 0.135** | 0.0533 | -0.305*** |
| | (0.0212) | (0.0207) | (0.0299) | (0.0219) | (0.0261) | (0.0490) | (0.0215) | (0.0640) | (0.0783) | (0.100) |
| More than 3.000 | 0.118*** | 0.157*** | 0.302*** | 0.131*** | 0.199*** | -0.0254 | 0.161*** | 0.0505 | 0.158 | -0.303* |
| | (0.0375) | (0.0366) | (0.0529) | (0.0387) | (0.0461) | (0.0867) | (0.0380) | (0.113) | (0.139) | (0.177) |
| Economic activity of household head (base cat: Govt. job) | | | | | | | | | | |
| Private sector job | -0.0197 | -0.0242 | -0.0539 | -0.0134 | 0.00269 | -0.169*** | -0.0177 | -0.179** | -0.181* | 0.0868 |
| | (0.0267) | (0.0260) | (0.0376) | (0.0275) | (0.0328) | (0.0616) | (0.0270) | (0.0804) | (0.0985) | (0.126) |
| Self employed | -0.0708*** | -0.0715*** | -0.0503 | -0.0432* | -0.114*** | -0.0190 | -0.00860 | -0.0945 | -0.225** | 0.335*** |
| | (0.0242) | (0.0236) | (0.0341) | (0.0249) | (0.0297) | (0.0559) | (0.0245) | (0.0729) | (0.0892) | (0.114) |
| Unemployed | -0.0821 | -0.000133 | -0.290** | -0.0507 | -0.0771 | 0.114 | -0.0869 | -0.290 | -1.004*** | -0.547 |
| | (0.0815) | (0.0796) | (0.115) | (0.0841) | (0.100) | (0.188) | (0.0826) | (0.246) | (0.301) | (0.385) |
| Economically inactive | -0.0282 | -0.0344 | 0.0177 | -0.0135 | -0.0833*** | -0.116** | -0.00232 | 0.00669 | -0.227** | -0.0828 |
| | (0.0251) | (0.0245) | (0.0354) | (0.0259) | (0.0308) | (0.0580) | (0.0254) | (0.0758) | (0.0927) | (0.118) |
| Constant | 5.242*** | 4.484*** | 4.922*** | 3.724*** | 4.440*** | 4.605*** | 3.404*** | 1.347*** | 1.776*** | 1.572*** |
| | (0.0887) | (0.0866) | (0.125) | (0.0915) | (0.109) | (0.205) | (0.0899) | (0.268) | (0.328) | (0.419) |
| Observations | 12,430 | 12,430 | 12,430 | 12,430 | 12,430 | 12,430 | 12,430 | 12,430 | 12,430 | 12,430 |
| R-squared | 0.290 | 0.504 | 0.380 | 0.277 | 0.308 | 0.181 | 0.332 | 0.233 | 0.382 | 0.204 |

Standard errors in parentheses

*** p<0.01, ** p<0.05, * p<0.1
We also control for the district fixed effect

Table A5: Sub-group wise expenditure OLS estimates for left-behind households (rural)

| Variables | (1) Cereals | (2) Pulses | (3) Milk and related products | (4) Oil, ghee and butter etc. | (5) Vegetables and fruits | (6) Egg, fish and meat | (7) Sugar | (8) Processed food | (9) Ready to eat | (10) Tobacco, liquor and intoxicants |
|---|---|---|---|---|---|---|---|---|---|---|
| | | | | | MPCE spendings on various sub-groups | | | | | |
| household with no emigrant (base cat.) | | | | | | | | | | |
| household with an emigrant | 0.115*** | 0.0579*** | 0.179*** | 0.0917*** | 0.177*** | 0.233*** | 0.0590*** | 0.359*** | -0.0199 | -0.329*** |
| | (0.0124) | (0.0104) | (0.0168) | (0.0122) | (0.0149) | (0.0309) | (0.0119) | (0.0453) | (0.0564) | (0.0729) |
| Dependency ratio | 0.000157*** | -7.90e-05 | 0.000378*** | 0.000104 | 0.000197** | 6.83e-05 | -7.73e-05 | 0.00123*** | -0.00106*** | -0.00131*** |
| | (7.59e-05) | (6.37e-05) | (0.000103) | (7.49e-05) | (9.13e-05) | (0.000190) | (7.30e-05) | (0.000278) | (0.000346) | (0.000447) |
| Gender of household head (base cat: male) | | | | | | | | | | |
| Female | -0.0399** | -0.00724 | -0.0447* | -0.0240 | 0.00333 | -0.0731 | -0.0126 | 0.00320 | -0.317*** | -0.643*** |
| | (0.0180) | (0.0151) | (0.0244) | (0.0177) | (0.0216) | (0.0449) | (0.0173) | (0.0659) | (0.0820) | (0.106) |
| Age of household head (base cat: 15-24 years) | | | | | | | | | | |
| 25-34 years | -0.232** | -0.157* | 0.0671 | -0.0226 | -0.123 | -0.0228 | -0.170* | -0.480 | -0.252 | -0.254 |
| | (0.102) | (0.0858) | (0.139) | (0.101) | (0.123) | (0.255) | (0.0984) | (0.375) | (0.467) | (0.603) |
| 35-44 years | -0.132 | -0.0924 | -0.00488 | 0.0434 | -0.0486 | 0.0384 | -0.0985 | -0.510 | -0.397 | -0.324 |
| | (0.101) | (0.0844) | (0.136) | (0.0994) | (0.121) | (0.251) | (0.0968) | (0.369) | (0.459) | (0.593) |
| 45-59 years | -0.111 | -0.0720 | 0.0599 | 0.103 | -0.0213 | 0.0236 | -0.0977 | -0.554 | -0.362 | -0.327 |
| | (0.100) | (0.0842) | (0.136) | (0.0991) | (0.121) | (0.251) | (0.0965) | (0.368) | (0.458) | (0.591) |
| 60 and above years | -0.106 | -0.0589 | 0.0816 | 0.105 | -0.000963 | -0.0592 | -0.0895 | -0.643* | -0.564 | -0.419 |
| | (0.101) | (0.0844) | (0.136) | (0.0993) | (0.121) | (0.251) | (0.0968) | (0.369) | (0.459) | (0.593) |
| Religion of household head (base cat: Hindu) | | | | | | | | | | |
| Christian | 0.0746*** | 0.0275*** | 0.0747*** | 0.0532*** | -0.0186 | 0.515*** | 0.0212* | 0.0637 | 0.0766 | -0.446*** |
| | (0.0121) | (0.0102) | (0.0164) | (0.0120) | (0.0146) | (0.0302) | (0.0116) | (0.0444) | (0.0552) | (0.0713) |
| Muslim | 0.125*** | 0.0213* | 0.0541*** | 0.0871*** | -0.0136 | 0.501*** | 0.102*** | 0.111** | 0.0772 | -1.222*** |
| | (0.0130) | (0.0109) | (0.0177) | (0.0129) | (0.0157) | (0.0326) | (0.0126) | (0.0479) | (0.0596) | (0.0769) |
| Education level of household head (base cat: Primary) | | | | | | | | | | |
| Upper Primary | 0.0370*** | 0.0445*** | 0.0637*** | 0.0161 | 0.0129 | -0.00861 | 0.00236 | 0.146*** | -0.00195 | -0.219*** |
| | (0.0132) | (0.0111) | (0.0180) | (0.0131) | (0.0159) | (0.0331) | (0.0128) | (0.0486) | (0.0605) | (0.0781) |
| Secondary | 0.104*** | 0.0710*** | 0.130*** | 0.0717*** | 0.0940*** | -0.0470 | 0.0320** | 0.267*** | -0.0737 | -0.571*** |
|  |  | (0.0132) | (0.0111) | (0.0179) | (0.0130) | (0.0159) | (0.0330) | (0.0127) | (0.0484) | (0.0602) | (0.0778) |
| Higher Secondary |  | 0.161*** | 0.0907*** | 0.263*** | 0.103*** | 0.189*** | -0.165*** | 0.0359 | 0.464*** | -0.0694 | -1.075*** |
|  |  | (0.0240) | (0.0202) | (0.0326) | (0.0237) | (0.0289) | (0.0601) | (0.0231) | (0.0882) | (0.110) | (0.142) |
| Degree and above |  | 0.247*** | 0.139*** | 0.450*** | 0.150*** | 0.248*** | -0.225*** | 0.142*** | 0.705*** | 0.269** | -1.438*** |
|  |  | (0.0231) | (0.0194) | (0.0314) | (0.0229) | (0.0278) | (0.0578) | (0.0223) | (0.0849) | (0.106) | (0.136) |
| not educated |  | -0.0938*** | -0.0433*** | -0.114*** | -0.0689*** | -0.133*** | -0.0536 | -0.0944*** | -0.0424 | 0.00747 | 0.574*** |
|  |  | (0.0173) | (0.0145) | (0.0234) | (0.0171) | (0.0208) | (0.0432) | (0.0166) | (0.0634) | (0.0789) | (0.102) |
| Literate without school education |  | 0.0207 | 0.0270 | -0.00591 | -0.132*** | -0.00324 | -0.0330 | 0.0659** | -0.115 | 0.256* | 0.229 |
|  |  | (0.0332) | (0.0279) | (0.0450) | (0.0328) | (0.0399) | (0.0830) | (0.0320) | (0.122) | (0.152) | (0.196) |
| Marital status of household head (base cat: Unmarried) |  |  |  |  |  |  |  |  |  |  |  |
| Married |  | 0.182*** | 0.0379 | -0.00492 | 0.158*** | 0.228*** | 0.239** | 0.0299 | 0.543*** | -0.448** | -0.152 |
|  |  | (0.0425) | (0.0357) | (0.0577) | (0.0420) | (0.0512) | (0.106) | (0.0409) | (0.156) | (0.194) | (0.251) |
| Divorced and others |  | 0.136*** | 0.0231 | -0.0144 | 0.128*** | 0.160*** | 0.205* | 0.0155 | 0.544*** | -0.289 | -0.322 |
|  |  | (0.0433) | (0.0364) | (0.0588) | (0.0428) | (0.0521) | (0.108) | (0.0417) | (0.159) | (0.198) | (0.255) |
| Size of household (base cat: 0-2) |  |  |  |  |  |  |  |  |  |  |  |
| 3 - 5 members |  | 0.614*** | 0.209*** | 0.329*** | 0.451*** | 0.535*** | 0.743*** | 0.464*** | 1.147*** | 0.758*** | 0.673*** |
|  |  | (0.0130) | (0.0109) | (0.0176) | (0.0129) | (0.0157) | (0.0325) | (0.0125) | (0.0477) | (0.0594) | (0.0767) |
| 6-10 members |  | 1.167*** | 0.345*** | 0.517*** | 0.700*** | 0.760*** | 0.993*** | 0.721*** | 1.658*** | 1.213*** | 1.349*** |
|  |  | (0.0161) | (0.0135) | (0.0218) | (0.0159) | (0.0194) | (0.0402) | (0.0155) | (0.0590) | (0.0735) | (0.0949) |
| 10 and above members |  | 1.636*** | 0.630*** | 0.777*** | 1.069*** | 1.078*** | 1.461*** | 1.107*** | 2.190*** | 1.595*** | 2.459*** |
|  |  | (0.0423) | (0.0355) | (0.0573) | (0.0418) | (0.0508) | (0.106) | (0.0407) | (0.155) | (0.193) | (0.249) |
| Land holdings in acres (base cat: no land) |  |  |  |  |  |  |  |  |  |  |  |
| 0.01-0.100 |  | 0.0218 | 0.0628*** | 0.125*** | 0.0801*** | 0.141*** | 0.0134 | 0.0527*** | 0.184*** | 0.256*** | 0.127 |
|  |  | (0.0188) | (0.0158) | (0.0255) | (0.0186) | (0.0226) | (0.0470) | (0.0181) | (0.0690) | (0.0859) | (0.111) |
| 0.101-0..200 |  | 0.0503** | 0.0657*** | 0.0719** | 0.0828*** | 0.0868*** | -0.0435 | 0.0582** | 0.0414 | 0.181* | 0.173 |
|  |  | (0.0237) | (0.0199) | (0.0321) | (0.0234) | (0.0285) | (0.0592) | (0.0228) | (0.0868) | (0.108) | (0.140) |
| 0.201-0.300 |  | 0.0667** | 0.0544** | 0.119*** | 0.121*** | 0.116*** | -0.102 | 0.0627** | -0.162* | -0.163 | 0.0247 |
|  |  | (0.0260) | (0.0218) | (0.0353) | (0.0257) | (0.0313) | (0.0650) | (0.0251) | (0.0955) | (0.119) | (0.153) |
| 0.301-.0.500 |  | 0.0959*** | 0.0772*** | 0.163*** | 0.0991*** | 0.129*** | -0.108** | 0.0385* | 0.00722 | 0.0186 | -0.237* |
|  |  | (0.0220) | (0.0185) | (0.0298) | (0.0217) | (0.0265) | (0.0549) | (0.0212) | (0.0806) | (0.100) | (0.130) |

| | | | | | | | | | | |
|---|---|---|---|---|---|---|---|---|---|---|
| 0.501-1.000 | 0.133*** | 0.0977*** | 0.268*** | 0.172*** | 0.227*** | -0.0981* | 0.121*** | 0.0870 | -0.0520 | -0.420*** |
| | (0.0211) | (0.0177) | (0.0286) | (0.0209) | (0.0254) | (0.0527) | (0.0203) | (0.0774) | (0.0964) | (0.124) |
| 1.001-3.000 | 0.137*** | 0.133*** | 0.231*** | 0.122*** | 0.152*** | -0.113* | 0.111*** | 0.127 | 0.117 | -0.429*** |
| | (0.0234) | (0.0196) | (0.0317) | (0.0231) | (0.0281) | (0.0584) | (0.0225) | (0.0858) | (0.107) | (0.138) |
| More than 3.000 | 0.124*** | 0.199*** | 0.317*** | 0.121*** | 0.189*** | -0.154 | 0.170*** | -0.00693 | 0.138 | -0.407* |
| | (0.0413) | (0.0346) | (0.0560) | (0.0408) | (0.0496) | (0.103) | (0.0397) | (0.151) | (0.188) | (0.243) |
| Economic activity of household head (base cat: Govt. job) | | | | | | | | | | |
| Private sector job | -0.0124 | -0.0632** | -0.0542 | -0.0103 | -0.0256 | -0.140* | -0.0467 | -0.208* | -0.257* | 0.327 |
| | (0.0341) | (0.0286) | (0.0462) | (0.0337) | (0.0410) | (0.0852) | (0.0328) | (0.125) | (0.156) | (0.201) |
| Self employed | -0.0351 | -0.0797*** | -0.0523 | -0.0346 | -0.135*** | 0.0144 | 0.00332 | -0.127 | -0.255* | 0.581*** |
| | (0.0308) | (0.0258) | (0.0417) | (0.0304) | (0.0370) | (0.0769) | (0.0296) | (0.113) | (0.140) | (0.181) |
| Unemployed | 0.0210 | 0.0336 | -0.324** | -0.0466 | -0.0664 | 0.202 | 0.0185 | -0.207 | -1.112** | -0.197 |
| | (0.0989) | (0.0830) | (0.134) | (0.0977) | (0.119) | (0.247) | (0.0952) | (0.363) | (0.451) | (0.583) |
| Economically inactive | 0.00135 | -0.0460* | 0.0188 | -0.00923 | -0.109*** | -0.0522 | -0.00116 | -0.0291 | -0.211 | 0.113 |
| | (0.0319) | (0.0267) | (0.0432) | (0.0315) | (0.0383) | (0.0796) | (0.0307) | (0.117) | (0.146) | (0.188) |
| Constant | 5.728*** | 4.987*** | 5.381*** | 4.099*** | 5.125*** | 5.317*** | 3.733*** | 1.299*** | 1.261** | 1.488** |
| | (0.109) | (0.0911) | (0.147) | (0.107) | (0.131) | (0.271) | (0.104) | (0.398) | (0.495) | (0.640) |
| | | | | | | | | | | |
| Observations | 9,668 | 9,668 | 9,668 | 9,668 | 9,668 | 9,668 | 9,668 | 9,668 | 9,668 | 9,668 |
| R-squared | 0.511 | 0.357 | 0.301 | 0.322 | 0.321 | 0.210 | 0.414 | 0.281 | 0.417 | 0.212 |

Standard errors in parentheses

*** p<0.01, ** p<0.05, * p<0.1

We also control for the district fixed effect

Table A6: Sub-group wise expenditure OLS estimates for left-behind households (urban)

| | MPCE spendings on various sub-groups | | | | | | | | | |
|---|---|---|---|---|---|---|---|---|---|---|
| | (1) | (2) | (3) | (4) | (5) | (6) | (7) | (8) | (9) | (10) |
| Variables | Cereals | Pulses | Milk and related products | Oil, ghee and butter etc. | Vegetables and fruits | Egg, fish and meat | Sugar | Processed food | Ready to eat | Tobacco, liquor and intoxicants |
| household with no emigrant (base cat.) | | | | | | | | | | |
| household with an emigrant | 0.0575*** | 0.0120 | 0.213*** | 0.102*** | 0.0957*** | 0.143** | 0.0279 | 0.443*** | -0.0294 | -0.333** |
| | (0.0211) | (0.0224) | (0.0302) | (0.0220) | (0.0260) | (0.0680) | (0.0232) | (0.0859) | (0.112) | (0.130) |
| Dependency ratio | 3.44e-05 | 5.65e-05 | 0.000516*** | -2.29e-05 | 0.000241 | 6.79e-05 | 0.000318** | 0.00197*** | -0.000490 | -0.00168** |
| | (0.000139) | (0.000148) | (0.000199) | (0.000145) | (0.000171) | (0.000448) | (0.000153) | (0.000566) | (0.000736) | (0.000854) |
| Gender of household head (base cat: male) | | | | | | | | | | |
| Female | -0.00683 | -0.0417 | -0.0801* | -0.0430 | 0.0128 | 0.136 | -0.0948*** | -0.0646 | -0.244 | -0.640*** |
| | (0.0316) | (0.0336) | (0.0451) | (0.0328) | (0.0389) | (0.102) | (0.0347) | (0.128) | (0.167) | (0.194) |
| Age of household head (base cat: 15-24 years) | | | | | | | | | | |
| 25-34 years | -0.0875 | -0.0416 | 0.0558 | 0.0350 | 0.562*** | 0.333 | -0.207 | 0.321 | 0.659 | 0.618 |
| | (0.160) | (0.170) | (0.229) | (0.167) | (0.197) | (0.516) | (0.176) | (0.652) | (0.848) | (0.984) |
| 35-44 years | 0.00866 | 0.00287 | 0.00837 | 0.0639 | 0.603*** | 0.316 | -0.121 | 0.146 | 0.379 | 0.386 |
| | (0.155) | (0.165) | (0.221) | (0.161) | (0.191) | (0.499) | (0.170) | (0.630) | (0.820) | (0.951) |
| 45-59 years | 0.0693 | 0.0330 | 0.102 | 0.181 | 0.665*** | 0.377 | -0.0849 | 0.304 | 0.412 | 0.205 |
| | (0.154) | (0.163) | (0.220) | (0.160) | (0.189) | (0.496) | (0.169) | (0.626) | (0.814) | (0.945) |
| 60 and above years | 0.0784 | 0.0125 | 0.0596 | 0.145 | 0.655*** | 0.253 | -0.0939 | 0.0898 | 0.171 | -0.0405 |
| | (0.154) | (0.164) | (0.220) | (0.160) | (0.189) | (0.496) | (0.169) | (0.626) | (0.815) | (0.945) |
| Religion of household head (base cat: Hindu) | | | | | | | | | | |
| Christian | 0.0146 | 0.0227 | 0.0784** | 0.0555** | -0.00642 | 0.653*** | 0.111*** | -0.127 | 0.244** | 0.0919 |
| | (0.0215) | (0.0228) | (0.0307) | (0.0223) | (0.0264) | (0.0691) | (0.0236) | (0.0872) | (0.114) | (0.132) |
| Muslim | 0.118*** | 0.0757*** | -0.0414 | -0.0968*** | 0.0553** | 0.504*** | 0.0321 | -0.132 | -0.217* | -1.080*** |
| | (0.0229) | (0.0244) | (0.0328) | (0.0238) | (0.0282) | (0.0738) | (0.0252) | (0.0932) | (0.121) | (0.141) |
| Education level of household head (base cat: Primary) | | | | | | | | | | |
| Upper Primary | -0.0177 | 0.00764 | 0.129*** | -0.0668** | 0.0318 | -0.0997 | -0.00897 | 0.190* | -0.249* | -0.664*** |
|  | (0.0263) | (0.0279) | (0.0376) | (0.0273) | (0.0324) | (0.0847) | (0.0289) | (0.107) | (0.139) | (0.161) |
| Secondary | 0.0466* | 0.0991*** | 0.228*** | 0.0346 | 0.0949*** | -0.204*** | 0.0154 | 0.376*** | -0.0856 | -0.832*** |
|  | (0.0244) | (0.0259) | (0.0349) | (0.0254) | (0.0300) | (0.0786) | (0.0268) | (0.0993) | (0.129) | (0.150) |
| Higher Secondary | 0.0192 | 0.112** | 0.364*** | 0.0845* | 0.178*** | -0.279** | 0.0240 | 0.698*** | 0.331 | -1.203*** |
|  | (0.0432) | (0.0459) | (0.0617) | (0.0449) | (0.0531) | (0.139) | (0.0474) | (0.175) | (0.228) | (0.265) |
| Degree and above | 0.121*** | 0.192*** | 0.531*** | 0.0916*** | 0.326*** | -0.495*** | 0.0751** | 0.772*** | 0.212 | -1.636*** |
|  | (0.0317) | (0.0336) | (0.0453) | (0.0329) | (0.0390) | (0.102) | (0.0348) | (0.129) | (0.168) | (0.194) |
| not educated | -0.101*** | -0.0517 | -0.0478 | -0.109*** | -0.158*** | 0.0248 | -0.184*** | -0.418*** | 0.131 | 0.759*** |
|  | (0.0363) | (0.0385) | (0.0518) | (0.0377) | (0.0446) | (0.117) | (0.0398) | (0.147) | (0.192) | (0.223) |
| Literate without school education | -0.177*** | 0.0738 | -0.0109 | 0.0136 | -0.102 | 0.0333 | -0.0729 | 0.147 | 0.927*** | 0.321 |
|  | (0.0662) | (0.0702) | (0.0945) | (0.0687) | (0.0813) | (0.213) | (0.0726) | (0.269) | (0.350) | (0.406) |
| Marital status of household head (base cat: Unmarried) | | | | | | | | | | |
| Married | 0.352*** | 0.0241 | -0.0768 | 0.153** | 0.839*** | 0.694*** | 0.117 | 0.685** | -0.766** | -0.0669 |
|  | (0.0697) | (0.0740) | (0.0995) | (0.0724) | (0.0856) | (0.224) | (0.0764) | (0.283) | (0.368) | (0.427) |
| Divorced and others | 0.248*** | 0.0564 | -0.0228 | 0.108 | 0.790*** | 0.464** | 0.196** | 0.890*** | -0.572 | -0.169 |
|  | (0.0715) | (0.0759) | (0.102) | (0.0743) | (0.0879) | (0.230) | (0.0785) | (0.291) | (0.378) | (0.439) |
| Size of household (base cat: 0-2) | | | | | | | | | | |
| 3 - 5 members | 0.582*** | 0.262*** | 0.363*** | 0.401*** | 0.479*** | 0.906*** | 0.427*** | 1.071*** | 0.888*** | 0.520*** |
|  | (0.0228) | (0.0242) | (0.0325) | (0.0237) | (0.0280) | (0.0733) | (0.0250) | (0.0925) | (0.120) | (0.140) |
| 6-10 members | 1.096*** | 0.393*** | 0.670*** | 0.641*** | 0.673*** | 1.132*** | 0.680*** | 1.586*** | 1.148*** | 1.207*** |
|  | (0.0286) | (0.0303) | (0.0408) | (0.0297) | (0.0351) | (0.0919) | (0.0313) | (0.116) | (0.151) | (0.175) |
| 10 and above members | 1.656*** | 0.690*** | 0.946*** | 0.796*** | 1.061*** | 1.650*** | 1.205*** | 2.228*** | 1.477*** | 1.937*** |
|  | (0.0853) | (0.0905) | (0.122) | (0.0886) | (0.105) | (0.274) | (0.0936) | (0.346) | (0.451) | (0.523) |
| Land holdings in acres (base cat: no land) | | | | | | | | | | |
| 0.01-0.100 | -0.00252 | 0.126*** | 0.135*** | 0.0280 | 0.108*** | 0.134 | 0.101*** | 0.292** | 0.329* | 0.0168 |
|  | (0.0325) | (0.0345) | (0.0464) | (0.0337) | (0.0399) | (0.104) | (0.0356) | (0.132) | (0.172) | (0.199) |
| 0.101-0..200 | 0.0941** | 0.159*** | 0.189*** | 0.128*** | 0.171*** | 0.306** | 0.190*** | 0.223 | 0.326 | 0.0879 |
|  | (0.0439) | (0.0466) | (0.0626) | (0.0456) | (0.0539) | (0.141) | (0.0481) | (0.178) | (0.232) | (0.269) |
| 0.201-0.300 | 0.0518 | 0.0741 | 0.168** | 0.0745 | 0.185*** | 0.0367 | 0.106* | 0.520** | 0.540* | 0.0148 |
|  | (0.0530) | (0.0563) | (0.0757) | (0.0551) | (0.0652) | (0.171) | (0.0582) | (0.215) | (0.280) | (0.325) |
| 0.301-.0.500 | 0.112** | 0.0577 | 0.196*** | 0.0819 | 0.273*** | 0.128 | 0.141** | 0.0676 | -0.143 | -0.280 |
| | (0.0529) | (0.0562) | (0.0756) | (0.0550) | (0.0651) | (0.170) | (0.0581) | (0.215) | (0.280) | (0.325) |
| 0.501-1.000 | 0.0581 | 0.119** | 0.184** | 0.0633 | 0.219*** | 0.292* | 0.127** | -0.0222 | -0.381 | -0.475 |
| | (0.0510) | (0.0541) | (0.0728) | (0.0529) | (0.0627) | (0.164) | (0.0559) | (0.207) | (0.270) | (0.313) |
| 1.001-3.000 | 0.217*** | 0.184** | 0.424*** | 0.178** | 0.362*** | 0.389* | 0.139* | 0.808*** | 0.364 | -0.0599 |
| | (0.0731) | (0.0776) | (0.104) | (0.0759) | (0.0898) | (0.235) | (0.0802) | (0.297) | (0.386) | (0.448) |
| More than 3.000 | 0.0792 | -0.0311 | 0.272 | 0.192 | 0.356** | 0.149 | 0.102 | 0.309 | 1.198* | -0.652 |
| | (0.123) | (0.131) | (0.176) | (0.128) | (0.151) | (0.396) | (0.135) | (0.500) | (0.650) | (0.754) |
| Economic activity of household head (base cat: Govt. job) | | | | | | | | | | |
| Private sector job | 0.00863 | 0.0747 | -0.0116 | 0.0262 | 0.0663 | -0.242* | 0.0732 | -0.366** | -0.359 | -0.292 |
| | (0.0457) | (0.0485) | (0.0652) | (0.0474) | (0.0561) | (0.147) | (0.0501) | (0.186) | (0.242) | (0.280) |
| Self employed | -0.0939** | 0.00803 | -0.00987 | -0.0315 | -0.0304 | 0.00687 | 0.0157 | -0.170 | -0.342 | 0.245 |
| | (0.0422) | (0.0448) | (0.0602) | (0.0438) | (0.0518) | (0.136) | (0.0463) | (0.171) | (0.223) | (0.259) |
| Unemployed | -0.0914 | 0.112 | 0.0604 | 0.165 | 0.205 | 0.288 | -0.127 | -0.521 | -1.770** | -1.725* |
| | (0.159) | (0.169) | (0.227) | (0.165) | (0.195) | (0.512) | (0.174) | (0.646) | (0.841) | (0.975) |
| Economically inactive | -0.0195 | 0.0632 | 0.0733 | 0.0427 | 0.0118 | -0.189 | 0.0576 | 0.00340 | -0.449* | -0.433 |
| | (0.0441) | (0.0468) | (0.0630) | (0.0458) | (0.0542) | (0.142) | (0.0484) | (0.179) | (0.233) | (0.271) |
| Constant | 5.666*** | 4.883*** | 5.417*** | 4.398*** | 4.139*** | 4.475*** | 4.056*** | 1.341* | 2.935*** | 2.150** |
| | (0.170) | (0.180) | (0.242) | (0.176) | (0.208) | (0.545) | (0.186) | (0.689) | (0.897) | (1.040) |
| | | | | | | | | | | |
| Observations | 2,762 | 2,762 | 2,762 | 2,762 | 2,762 | 2,762 | 2,762 | 2,762 | 2,762 | 2,762 |
| R-squared | 0.510 | 0.299 | 0.338 | 0.356 | 0.439 | 0.228 | 0.402 | 0.299 | 0.383 | 0.238 |

Standard errors in parentheses

*** p<0.01, ** p<0.05, * p<0.1

We also control for the district fixed effect

| | (1) | (2) | (3) | (4) | (5) | (6) | (7) | (8) | (9) | (10) |
|---|---|---|---|---|---|---|---|---|---|---|
| Variables | Cereals | Pulses | Milk and related products | Oil, ghee and butter etc. | Vegetables and fruits | Egg, fish and meat | Sugar | Processed food | Ready to eat | Tobacco, liquor and intoxicants |
| household with no emigrant (base cat.) | | | | | | | | | | |
| household with an emigrant | 0.248*** | 0.429*** | 0.526*** | 0.116 | 0.314*** | 0.178 | 0.162* | 0.864*** | 1.251*** | -0.217 |
| | (0.0790) | (0.0919) | (0.0974) | (0.0886) | (0.0822) | (0.126) | (0.0841) | (0.207) | (0.284) | (0.259) |
| Dependency ratio | -0.000602*** | -0.00100*** | -0.000505*** | -0.000551*** | -0.000551*** | -0.000462*** | -0.000705*** | 6.62e-05 | -0.00209*** | -0.00123*** |
| | (8.60e-05) | (0.000100) | (0.000119) | (9.74e-05) | (0.000105) | (0.000157) | (9.53e-05) | (0.000256) | (0.000341) | (0.000355) |
| Gender of household head (base cat: male) | | | | | | | | | | |
| Female | -0.0162 | -0.0774** | -0.106*** | 0.0254 | 0.0137 | 0.0509 | -0.00892 | -0.129 | -0.661*** | -0.543*** |
| | (0.0329) | (0.0352) | (0.0382) | (0.0365) | (0.0375) | (0.0566) | (0.0348) | (0.0847) | (0.129) | (0.113) |
| Age of household head (base cat: 15-24 years) | | | | | | | | | | |
| 25-34 years | -0.0863 | -0.0189 | 0.182 | 0.0993 | 0.156 | 0.179 | -0.0666 | -0.0743 | 0.132 | 0.0476 |
| | (0.0856) | (0.0714) | (0.144) | (0.103) | (0.204) | (0.243) | (0.0905) | (0.271) | (0.376) | (0.363) |
| 35-44 years | -0.0587 | -0.0328 | 0.0461 | 0.0944 | 0.151 | 0.165 | -0.0631 | -0.162 | -0.0613 | -0.0504 |
| | (0.0863) | (0.0695) | (0.143) | (0.100) | (0.204) | (0.235) | (0.0905) | (0.262) | (0.366) | (0.359) |
| 45-59 years | -0.0443 | -0.0436 | 0.0820 | 0.161 | 0.168 | 0.175 | -0.0665 | -0.211 | -0.130 | -0.0682 |
| | (0.0868) | (0.0681) | (0.143) | (0.101) | (0.204) | (0.236) | (0.0908) | (0.265) | (0.365) | (0.360) |
| 60 and above years | -0.0383 | -0.0320 | 0.0871 | 0.155 | 0.187 | 0.0782 | -0.0598 | -0.301 | -0.281 | -0.184 |
| | (0.0863) | (0.0674) | (0.145) | (0.102) | (0.204) | (0.239) | (0.0908) | (0.266) | (0.365) | (0.361) |
| Religion of household head (base cat: Hindu) | | | | | | | | | | |
| Christian | 0.0591*** | 0.00905 | 0.0573** | 0.0604*** | -0.0244 | 0.492*** | 0.0395* | -0.0203 | 0.0165 | -0.245*** |
| | (0.0184) | (0.0210) | (0.0231) | (0.0223) | (0.0264) | (0.0442) | (0.0215) | (0.0535) | (0.0745) | (0.0728) |
| Muslim | 0.0502* | -0.103*** | -0.0815** | -0.00639 | -0.0808** | 0.411*** | 0.00772 | -0.108 | -0.327*** | -0.913*** |
| | (0.0278) | (0.0334) | (0.0395) | (0.0379) | (0.0321) | (0.0506) | (0.0302) | (0.0802) | (0.116) | (0.102) |
| Education level of household head (base cat: Primary) | | | | | | | | | | |
| Upper Primary | 0.0389*** | 0.0457*** | 0.0747*** | 0.00871 | 0.0238 | -0.0115 | 0.00772 | 0.122*** | -0.0620 | -0.232*** |
| | (0.0129) | (0.0125) | (0.0186) | (0.0142) | (0.0159) | (0.0245) | (0.0123) | (0.0402) | (0.0485) | (0.0589) |
| Secondary | 0.113*** | 0.0871*** | 0.145*** | 0.0794*** | 0.110*** | -0.0547* | 0.0418*** | 0.228*** | -0.120* | -0.476*** |
| | (0.0159) | (0.0164) | (0.0188) | (0.0165) | (0.0194) | (0.0301) | (0.0155) | (0.0487) | (0.0622) | (0.0584) |
| Higher Secondary | 0.173*** | 0.120*** | 0.294*** | 0.144*** | 0.230*** | -0.113** | 0.0741*** | 0.413*** | -0.0535 | -0.849*** |
| | (0.0255) | (0.0292) | (0.0337) | (0.0254) | (0.0289) | (0.0521) | (0.0275) | (0.0751) | (0.101) | (0.0902) |
| Degree and above | 0.242*** | 0.186*** | 0.476*** | 0.167*** | 0.319*** | -0.232*** | 0.155*** | 0.636*** | 0.193* | -1.117*** |

Table A7: Sub-group wise expenditure IV estimates for left-behind households (rural+urban)
MPCE spendings on various sub-groups

|  | (0.0228) | (0.0301) | (0.0310) | (0.0228) | (0.0294) | (0.0769) | (0.0302) | (0.0733) | (0.103) | (0.0915) |
|---|---|---|---|---|---|---|---|---|---|---|
| not educated | -0.0966*** | -0.0360* | -0.0860*** | -0.0797*** | -0.135*** | -0.0418 | -0.106*** | -0.0750 | 0.0814 | 0.473*** |
|  | (0.0200) | (0.0195) | (0.0268) | (0.0237) | (0.0265) | (0.0421) | (0.0215) | (0.0556) | (0.0712) | (0.0940) |
| Literate without school education | -0.0340 | 0.00847 | -0.0130 | -0.0855 | -0.0299 | -0.0115 | 0.0431 | -0.0567 | 0.222 | 0.254 |
|  | (0.0446) | (0.0395) | (0.0585) | (0.0531) | (0.0527) | (0.0879) | (0.0417) | (0.106) | (0.182) | (0.173) |
| Marital status of household head (base cat: Unmarried) | | | | | | | | | | |
| Married | 0.0739* | -0.174*** | -0.213*** | 0.0506 | 0.228** | 0.247** | -0.0764* | 0.288* | -0.847*** | -0.191 |
|  | (0.0418) | (0.0491) | (0.0565) | (0.0489) | (0.104) | (0.121) | (0.0458) | (0.161) | (0.194) | (0.221) |
| Divorced and others | 0.0666* | -0.0459 | -0.0872 | 0.0288 | 0.220** | 0.161 | -0.0275 | 0.479*** | -0.236 | -0.264 |
|  | (0.0357) | (0.0448) | (0.0539) | (0.0429) | (0.0985) | (0.109) | (0.0395) | (0.152) | (0.174) | (0.206) |
| Size of household (base cat: 0-2) | | | | | | | | | | |
| 3 - 5 members | -0.154*** | -0.507*** | -0.406*** | -0.328*** | -0.239*** | -0.00167 | -0.292*** | 0.428*** | 0.449*** | 0.295*** |
|  | (0.0163) | (0.0177) | (0.0212) | (0.0184) | (0.0212) | (0.0378) | (0.0170) | (0.0595) | (0.0637) | (0.0620) |
| 6-10 members | -0.0979*** | -0.839*** | -0.676*** | -0.569*** | -0.505*** | -0.239*** | -0.512*** | 0.424*** | 0.579*** | 0.612*** |
|  | (0.0224) | (0.0253) | (0.0298) | (0.0246) | (0.0275) | (0.0477) | (0.0226) | (0.0723) | (0.0787) | (0.0779) |
| 10 and above members | -0.212*** | -1.168*** | -1.030*** | -0.795*** | -0.768*** | -0.352*** | -0.665*** | 0.316*** | 0.457*** | 1.045*** |
|  | (0.0372) | (0.0622) | (0.0584) | (0.0536) | (0.0548) | (0.0635) | (0.0565) | (0.118) | (0.160) | (0.174) |
| Land holdings in acres (base cat: no land) | | | | | | | | | | |
| 0.01-0.100 | -0.0208 | 0.0307 | 0.0876** | 0.0387 | 0.104*** | 0.0219 | 0.0387 | 0.138* | 0.144 | 0.0830 |
|  | (0.0295) | (0.0279) | (0.0370) | (0.0356) | (0.0325) | (0.0632) | (0.0281) | (0.0840) | (0.115) | (0.109) |
| 0.101-0..200 | 0.0323 | 0.0439 | 0.0593 | 0.0809** | 0.0835** | 0.0358 | 0.0700** | 0.0450 | 0.0771 | 0.114 |
|  | (0.0272) | (0.0273) | (0.0413) | (0.0321) | (0.0348) | (0.0651) | (0.0304) | (0.0937) | (0.125) | (0.127) |
| 0.201-0.300 | 0.0271 | -0.000314 | 0.0797** | 0.0992*** | 0.0987*** | -0.0493 | 0.0496* | -0.0720 | -0.125 | 0.0147 |
|  | (0.0267) | (0.0384) | (0.0383) | (0.0269) | (0.0332) | (0.0848) | (0.0285) | (0.0972) | (0.135) | (0.130) |
| 0.301-.0.500 | 0.0630*** | 0.0120 | 0.119*** | 0.0802** | 0.117*** | -0.0385 | 0.0276 | -0.0342 | -0.134 | -0.198* |
|  | (0.0243) | (0.0275) | (0.0360) | (0.0321) | (0.0354) | (0.0665) | (0.0312) | (0.0965) | (0.139) | (0.118) |
| 0.501-1.000 | 0.0920*** | 0.0442 | 0.214*** | 0.142*** | 0.196*** | -0.00926 | 0.0961*** | 0.0472 | -0.153 | -0.327*** |
|  | (0.0297) | (0.0301) | (0.0365) | (0.0327) | (0.0316) | (0.0832) | (0.0268) | (0.0931) | (0.137) | (0.119) |
| 1.001-3.000 | 0.113*** | 0.0765** | 0.202*** | 0.109*** | 0.141*** | -0.0256 | 0.0866*** | 0.0937 | -0.0415 | -0.305** |
|  | (0.0282) | (0.0325) | (0.0437) | (0.0389) | (0.0376) | (0.0881) | (0.0291) | (0.0936) | (0.129) | (0.119) |
| More than 3.000 | 0.104*** | 0.116** | 0.266*** | 0.131*** | 0.183*** | -0.0209 | 0.150*** | -0.0104 | 0.0168 | -0.304 |
|  | (0.0337) | (0.0461) | (0.0676) | (0.0503) | (0.0463) | (0.144) | (0.0409) | (0.110) | (0.149) | (0.202) |
| Economic activity of household head (base cat: Govt. job) | | | | | | | | | | |
| Private sector job | -0.0220 | -0.0308 | -0.0596 | -0.0135 | 0.000179 | -0.169** | -0.0194 | -0.188** | -0.204* | 0.0866 |
|  | (0.0264) | (0.0342) | (0.0416) | (0.0285) | (0.0367) | (0.0793) | (0.0398) | (0.0831) | (0.122) | (0.140) |
| Self employed | -0.0765*** | -0.0876*** | -0.0646* | -0.0434* | -0.121*** | -0.0173 | -0.0128 | -0.119* | -0.280*** | 0.335*** |
|  | (0.0235) | (0.0287) | (0.0353) | (0.0263) | (0.0338) | (0.0739) | (0.0288) | (0.0674) | (0.0979) | (0.125) |
| Unemployed | -0.115 | -0.0935 | -0.373** | -0.0523 | -0.113 | 0.124 | -0.111 | -0.429* | -1.326*** | -0.549* |
| | (0.0771) | (0.0856) | (0.156) | (0.103) | (0.0949) | (0.134) | (0.0882) | (0.231) | (0.339) | (0.298) |
| Economically inactive | -0.0515* | -0.101*** | -0.0407 | -0.0146 | -0.109*** | -0.109 | -0.0193 | -0.0917 | -0.455*** | -0.0840 |
| | (0.0292) | (0.0323) | (0.0405) | (0.0297) | (0.0364) | (0.0802) | (0.0339) | (0.0760) | (0.119) | (0.135) |
| Constant | 5.277*** | 4.582*** | 5.008*** | 3.725*** | 4.477*** | 4.594*** | 3.429*** | 1.492*** | 2.112*** | 1.573*** |
| | (0.102) | (0.0978) | (0.156) | (0.121) | (0.264) | (0.290) | (0.118) | (0.345) | (0.489) | (0.424) |
| Observations | 12,430 | 12,430 | 12,430 | 12,430 | 12,430 | 12,430 | 12,430 | 12,430 | 12,430 | 12,430 |
| R-squared | 0.281 | 0.451 | 0.356 | 0.277 | 0.302 | 0.181 | 0.327 | 0.214 | 0.327 | 0.204 |

Robust standard errors in parentheses

*** p<0.01, ** p<0.05, * p<0.1

We also control for the district fixed effect

Table A8: Sub-group wise expenditure IV estimates for left-behind households (rural)

| | (1) | (2) | (3) | (4) | (5) | (6) | (7) | (8) | (9) | (10) |
|---|---|---|---|---|---|---|---|---|---|---|
| | | | | | MPCE spendings on various sub-groups | | | | | |
| Variables | Cereals | Pulses | Milk and related products | Oil, ghee and butter etc. | Vegetables and fruits | Egg, fish and meat | Sugar | Processed food | Ready to eat | Tobacco, liquor and intoxicants |
| household with no emigrant (base cat.) | | | | | | | | | | |
| household with an emigrant | 0.323*** | 0.472*** | 0.505*** | 0.106 | 0.385*** | 0.274* | 0.235** | 0.944*** | 1.216*** | -0.745** |
| | (0.0853) | (0.103) | (0.108) | (0.0914) | (0.0937) | (0.150) | (0.0926) | (0.265) | (0.361) | (0.357) |
| Dependency ratio | -2.16e-05 | -0.000435*** | 9.71e-05 | 9.19e-05 | 1.81e-05 | 3.30e-05 | -0.000229** | 0.000723* | -0.00212*** | -0.000957* |
| | (9.87e-05) | (0.000108) | (0.000135) | (0.000106) | (0.000120) | (0.000195) | (0.000110) | (0.000374) | (0.000482) | (0.000513) |
| Gender of household head (base cat: male) | | | | | | | | | | |
| Female | -0.109*** | -0.145*** | -0.153*** | -0.0289 | -0.0657 | -0.0868 | -0.0713* | -0.191* | -0.728*** | -0.504*** |
| | (0.0348) | (0.0383) | (0.0430) | (0.0381) | (0.0404) | (0.0649) | (0.0381) | (0.107) | (0.151) | (0.157) |
| Age of household head (base cat: 15-24 years) | | | | | | | | | | |
| 25-34 years | -0.217** | -0.126* | 0.0917 | -0.0215 | -0.107 | -0.0197 | -0.156 | -0.436 | -0.159 | -0.285 |
| | (0.0930) | (0.0755) | (0.174) | (0.129) | (0.144) | (0.291) | (0.108) | (0.361) | (0.446) | (0.582) |
| 35-44 years | -0.122 | -0.0728 | 0.0106 | 0.0441 | -0.0387 | 0.0403 | -0.0901 | -0.482 | -0.338 | -0.344 |
| | (0.0947) | (0.0723) | (0.172) | (0.127) | (0.145) | (0.279) | (0.109) | (0.349) | (0.435) | (0.583) |
| 45-59 years | -0.120 | -0.0907 | 0.0452 | 0.102 | -0.0307 | 0.0218 | -0.106 | -0.580 | -0.418 | -0.308 |
| | (0.0950) | (0.0708) | (0.171) | (0.127) | (0.143) | (0.274) | (0.109) | (0.354) | (0.429) | (0.585) |
| 60 and above years | -0.115 | -0.0763 | 0.0680 | 0.104 | -0.00967 | -0.0609 | -0.0969 | -0.668* | -0.616 | -0.401 |
| | (0.0943) | (0.0705) | (0.174) | (0.129) | (0.144) | (0.284) | (0.109) | (0.354) | (0.433) | (0.588) |
| Religion of household head (base cat: Hindu) | | | | | | | | | | |
| Christian | 0.0633*** | 0.00510 | 0.0570** | 0.0524** | -0.0299 | 0.513*** | 0.0116 | 0.0320 | 0.00966 | -0.423*** |
| | (0.0204) | (0.0200) | (0.0256) | (0.0242) | (0.0289) | (0.0538) | (0.0209) | (0.0737) | (0.105) | (0.104) |
| Muslim | 0.0733** | -0.0810*** | -0.0265 | 0.0835** | -0.0649* | 0.490*** | 0.0586* | -0.0333 | -0.228 | -1.119*** |
| | (0.0289) | (0.0312) | (0.0453) | (0.0389) | (0.0381) | (0.0583) | (0.0342) | (0.101) | (0.147) | (0.149) |
| Education level of household head (base cat: Primary) | | | | | | | | | | |
| Upper Primary | 0.0312** | 0.0330*** | 0.0546** | 0.0157 | 0.00715 | -0.00975 | -0.00254 | 0.129** | -0.0363 | -0.208** |
| | (0.0159) | (0.0123) | (0.0213) | (0.0155) | (0.0189) | (0.0327) | (0.0148) | (0.0563) | (0.0674) | (0.0843) |
| Secondary | 0.0899*** | 0.0421*** | 0.108*** | 0.0707*** | 0.0795*** | -0.0499 | 0.0197 | 0.226*** | -0.160* | -0.542*** |
| | (0.0182) | (0.0159) | (0.0212) | (0.0184) | (0.0216) | (0.0350) | (0.0173) | (0.0684) | (0.0885) | (0.0838) |
| Higher Secondary | 0.142*** | 0.0526* | 0.233*** | 0.102*** | 0.170*** | -0.169** | 0.0196 | 0.410*** | -0.183 | -1.037*** |
| | (0.0283) | (0.0293) | (0.0381) | (0.0272) | (0.0309) | (0.0685) | (0.0291) | (0.0954) | (0.139) | (0.131) |
| Degree and above | 0.233*** | 0.110*** | 0.428*** | 0.149*** | 0.233*** | -0.228*** | 0.130*** | 0.664*** | 0.183 | -1.409*** |
| | (0.0264) | (0.0278) | (0.0361) | (0.0258) | (0.0346) | (0.0757) | (0.0287) | (0.105) | (0.144) | (0.143) |
| not educated | -0.0818*** | -0.0195 | -0.0948*** | -0.0680** | -0.121*** | -0.0512 | -0.0842*** | -0.00872 | 0.0785 | 0.550*** |
| | (0.0234) | (0.0194) | (0.0289) | (0.0268) | (0.0314) | (0.0541) | (0.0253) | (0.0736) | (0.101) | (0.133) |
| Literate without school education | 0.00872 | 0.00300 | -0.0248 | -0.133*** | -0.0153 | -0.0354 | 0.0556 | -0.148 | 0.185 | 0.253 |
| | (0.0445) | (0.0412) | (0.0554) | (0.0448) | (0.0583) | (0.110) | (0.0398) | (0.143) | (0.181) | (0.219) |
| Marital status of household head (base cat: Unmarried) | | | | | | | | | | |
| Married | 0.127*** | -0.0718 | -0.0914 | 0.154*** | 0.173* | 0.228* | -0.0168 | 0.389** | -0.776*** | -0.0415 |
| | (0.0478) | (0.0455) | (0.0634) | (0.0563) | (0.0885) | (0.138) | (0.0475) | (0.198) | (0.247) | (0.284) |
| Divorced and others | 0.152*** | 0.0552 | 0.0109 | 0.129** | 0.176* | 0.208 | 0.0292 | 0.589*** | -0.193 | -0.354 |
| | (0.0441) | (0.0395) | (0.0641) | (0.0505) | (0.0913) | (0.129) | (0.0420) | (0.187) | (0.223) | (0.264) |
| Size of household (base cat: 0-2) | | | | | | | | | | |
| 3 - 5 members | 0.634*** | 0.249*** | 0.361*** | 0.452*** | 0.555*** | 0.746*** | 0.481*** | 1.203*** | 0.877*** | 0.633*** |
| | (0.0200) | (0.0195) | (0.0244) | (0.0223) | (0.0265) | (0.0449) | (0.0198) | (0.0772) | (0.0835) | (0.0834) |
| 6-10 members | 1.192*** | 0.394*** | 0.556*** | 0.702*** | 0.785*** | 0.997*** | 0.742*** | 1.728*** | 1.361*** | 1.300*** |
| | (0.0273) | (0.0277) | (0.0328) | (0.0298) | (0.0335) | (0.0599) | (0.0270) | (0.0935) | (0.110) | (0.112) |
| 10 and above members | 1.630*** | 0.619*** | 0.768*** | 1.069*** | 1.072*** | 1.459*** | 1.102*** | 2.174*** | 1.561*** | 2.470*** |
| | (0.0470) | (0.0718) | (0.0676) | (0.0596) | (0.0658) | (0.0790) | (0.0643) | (0.178) | (0.294) | (0.333) |
| Land holdings in acres (base cat: no land) | | | | | | | | | | |
| 0.01-0.100 | 0.0125 | 0.0442 | 0.110*** | 0.0795* | 0.132*** | 0.0115 | 0.0448 | 0.158 | 0.201 | 0.146 |
| | (0.0369) | (0.0335) | (0.0427) | (0.0430) | (0.0384) | (0.0822) | (0.0315) | (0.131) | (0.186) | (0.181) |
| 0.101-0..200 | 0.0421 | 0.0493* | 0.0590 | 0.0822** | 0.0785** | -0.0451 | 0.0512* | 0.0182 | 0.132 | 0.189 |
| | (0.0310) | (0.0264) | (0.0482) | (0.0343) | (0.0382) | (0.0797) | (0.0296) | (0.141) | (0.185) | (0.195) |
| 0.201-0.300 | 0.0469 | 0.0149 | 0.0875** | 0.120*** | 0.0965** | -0.106 | 0.0459 | -0.218 | -0.280 | 0.0644 |
| | (0.0342) | (0.0465) | (0.0446) | (0.0307) | (0.0415) | (0.113) | (0.0313) | (0.150) | (0.208) | (0.204) |
| 0.301-.0.500 | 0.0771*** | 0.0398 | 0.133*** | 0.0978*** | 0.110*** | -0.112 | 0.0225 | -0.0456 | -0.0930 | -0.199 |
| | (0.0288) | (0.0295) | (0.0390) | (0.0333) | (0.0425) | (0.0857) | (0.0338) | (0.136) | (0.209) | (0.179) |
| 0.501-1.000 | 0.119*** | 0.0690** | 0.245*** | 0.171*** | 0.212*** | -0.101 | 0.109*** | 0.0464 | -0.138 | -0.391** |
| | (0.0323) | (0.0321) | (0.0368) | (0.0337) | (0.0373) | (0.107) | (0.0285) | (0.131) | (0.208) | (0.176) |
| 1.001-3.000 | 0.120*** | 0.0996*** | 0.205*** | 0.120*** | 0.135*** | -0.116 | 0.0972*** | 0.0801 | 0.0177 | -0.396** |
| | (0.0315) | (0.0328) | (0.0447) | (0.0404) | (0.0417) | (0.114) | (0.0298) | (0.132) | (0.187) | (0.171) |
| More than 3.000 | 0.104*** | 0.159*** | 0.286*** | 0.120** | 0.169*** | -0.158 | 0.153*** | -0.0636 | 0.0184 | -0.367 |
| | (0.0370) | (0.0516) | (0.0771) | (0.0606) | (0.0536) | (0.204) | (0.0456) | (0.166) | (0.208) | (0.292) |
| Economic activity of household head (base cat: Govt. job) | | | | | | | | | | |
| Private sector job | -0.0162 | -0.0708** | -0.0602 | -0.0106 | -0.0295 | -0.141 | -0.0500 | -0.219* | -0.280 | 0.335 |
| | (0.0304) | (0.0329) | (0.0531) | (0.0335) | (0.0427) | (0.107) | (0.0396) | (0.126) | (0.175) | (0.215) |
| Self employed | -0.0438 | -0.0971*** | -0.0660 | -0.0352 | -0.144*** | 0.0126 | -0.00407 | -0.152 | -0.307** | 0.599*** |
| | (0.0288) | (0.0335) | (0.0429) | (0.0321) | (0.0394) | (0.0928) | (0.0342) | (0.103) | (0.150) | (0.188) |
| Unemployed | -0.0450 | -0.0979 | -0.428** | -0.0512 | -0.132 | 0.189 | -0.0375 | -0.393 | -1.505*** | -0.0647 |
| | (0.108) | (0.111) | (0.216) | (0.129) | (0.113) | (0.161) | (0.115) | (0.325) | (0.566) | (0.508) |
| Economically inactive | -0.0363 | -0.121*** | -0.0403 | -0.0119 | -0.147*** | -0.0597 | -0.0331 | -0.135 | -0.435** | 0.188 |
| | (0.0337) | (0.0378) | (0.0484) | (0.0346) | (0.0426) | (0.101) | (0.0388) | (0.114) | (0.172) | (0.197) |
| Constant | 5.769*** | 5.069*** | 5.446*** | 4.102*** | 5.167*** | 5.326*** | 3.768*** | 1.415*** | 1.507*** | 1.406** |
| | (0.115) | (0.101) | (0.185) | (0.146) | (0.175) | (0.304) | (0.122) | (0.437) | (0.554) | (0.686) |
| Observations | 9,668 | 9,668 | 9,668 | 9,668 | 9,668 | 9,668 | 9,668 | 9,668 | 9,668 | 9,668 |
| R-squared | 0.497 | 0.250 | 0.273 | 0.322 | 0.307 | 0.210 | 0.400 | 0.269 | 0.388 | 0.209 |

Robust standard errors in parentheses

*** p<0.01, ** p<0.05, * p<0.1

We also control for the district fixed effect

Table A9: Sub-group wise expenditure IV estimates for left-behind households (urban)

| | MPCE spendings on various sub-groups | | | | | | | | | |
|---|---|---|---|---|---|---|---|---|---|---|
| | (1) | (2) | (3) | (4) | (5) | (6) | (7) | (8) | (9) | (10) |
| Variables | Cereals | Pulses | Milk and related products | Oil, ghee and butter etc. | Vegetables and fruits | Egg, fish and meat | Sugar | Processed food | Ready to eat | Tobacco, liquor and intoxicants |
| household with no emigrant (base cat.) | | | | | | | | | | |
| household with an emigrant | 0.0366 | 0.386** | 0.544*** | 0.257 | 0.0690 | 0.0810 | -0.105 | 1.656** | 3.234*** | 1.038 |
| | (0.181) | (0.182) | (0.172) | (0.240) | (0.175) | (0.339) | (0.200) | (0.742) | (1.217) | (0.856) |
| Dependency ratio | 5.11e-05 | -0.000244 | 0.000250 | -0.000148 | 0.000262 | 0.000118 | -0.000425** | 0.000996 | -0.00311*** | -0.00278*** |
| | (0.000185) | (0.000212) | (0.000228) | (0.000217) | (0.000207) | (0.000476) | (0.000200) | (0.000750) | (0.00116) | (0.000940) |
| Gender of household head (base cat: male) | | | | | | | | | | |
| Female | -0.000103 | -0.162** | -0.187*** | -0.0930 | 0.0214 | 0.156 | -0.0519 | -0.456 | -1.296** | -1.081*** |
| | (0.0729) | (0.0722) | (0.0654) | (0.0823) | (0.0760) | (0.147) | (0.0765) | (0.285) | (0.569) | (0.370) |
| Age of household head (base cat: 15-24 years) | | | | | | | | | | |
| 25-34 years | -0.0860 | -0.0682 | 0.0323 | 0.0239 | 0.563 | 0.338 | -0.198 | 0.234 | 0.426 | 0.521 |
| | (0.168) | (0.0769) | (0.153) | (0.143) | (0.629) | (0.607) | (0.135) | (0.680) | (1.050) | (0.981) |
| 35-44 years | 0.0107 | -0.0332 | -0.0235 | 0.0489 | 0.605 | 0.322 | -0.108 | 0.0288 | 0.0646 | 0.253 |
| | (0.154) | (0.0660) | (0.152) | (0.133) | (0.625) | (0.590) | (0.127) | (0.650) | (1.017) | (0.939) |
| 45-59 years | 0.0729 | -0.0326 | 0.0437 | 0.154 | 0.669 | 0.388 | -0.0616 | 0.0915 | -0.159 | -0.0352 |
| | (0.157) | (0.0694) | (0.152) | (0.137) | (0.633) | (0.601) | (0.129) | (0.663) | (1.050) | (0.943) |
| 60 and above years | 0.0825 | -0.0610 | -0.00532 | 0.114 | 0.660 | 0.265 | -0.0678 | -0.148 | -0.469 | -0.309 |
| | (0.161) | (0.0716) | (0.155) | (0.138) | (0.629) | (0.587) | (0.134) | (0.696) | (1.061) | (0.939) |
| Religion of household head (base cat: Hindu) | | | | | | | | | | |
| Christian | 0.0161 | -0.00389 | 0.0549 | 0.0445 | -0.00453 | 0.657*** | 0.121** | -0.213* | 0.0121 | -0.00537 |
| | (0.0423) | (0.0510) | (0.0377) | (0.0404) | (0.0434) | (0.110) | (0.0472) | (0.116) | (0.207) | (0.220) |
| Muslim | 0.123** | -0.0121 | -0.119* | -0.133** | 0.0616 | 0.519*** | 0.0632 | -0.416* | -0.982*** | -1.401*** |
| | (0.0588) | (0.0880) | (0.0638) | (0.0649) | (0.0490) | (0.139) | (0.0709) | (0.247) | (0.376) | (0.326) |
| Education level of household head (base cat: Primary) | | | | | | | | | | |
| Upper Primary | -0.0177 | 0.00910 | 0.130*** | -0.0662** | 0.0317 | -0.0999* | -0.00949 | 0.194* | -0.237 | -0.659*** |
|  | (0.0239) | (0.0326) | (0.0421) | (0.0310) | (0.0278) | (0.0526) | (0.0229) | (0.106) | (0.145) | (0.193) |
| Secondary | 0.0478* | 0.0774** | 0.209*** | 0.0256 | 0.0964*** | -0.201** | 0.0231 | 0.306** | -0.274* | -0.911*** |
|  | (0.0280) | (0.0348) | (0.0393) | (0.0273) | (0.0326) | (0.0900) | (0.0292) | (0.119) | (0.146) | (0.181) |
| Higher Secondary | 0.0215 | 0.0715 | 0.328*** | 0.0677 | 0.181*** | -0.272*** | 0.0384 | 0.567*** | -0.0215 | -1.351*** |
|  | (0.0364) | (0.0594) | (0.0650) | (0.0451) | (0.0535) | (0.105) | (0.0591) | (0.197) | (0.255) | (0.282) |
| Degree and above | 0.123*** | 0.158*** | 0.501*** | 0.0774** | 0.328*** | -0.489*** | 0.0873* | 0.660*** | -0.0874 | -1.762*** |
|  | (0.0439) | (0.0578) | (0.0600) | (0.0367) | (0.0461) | (0.178) | (0.0466) | (0.178) | (0.236) | (0.267) |
| not educated | -0.102** | -0.0340 | -0.0322 | -0.102** | -0.160*** | 0.0219 | -0.191*** | -0.361* | 0.286 | 0.823*** |
|  | (0.0437) | (0.0537) | (0.0594) | (0.0427) | (0.0406) | (0.129) | (0.0562) | (0.204) | (0.246) | (0.286) |
| Literate without school education | -0.177** | 0.0719 | -0.0126 | 0.0128 | -0.102 | 0.0336 | -0.0723 | 0.141 | 0.910* | 0.314 |
|  | (0.0855) | (0.0623) | (0.159) | (0.0863) | (0.0630) | (0.226) | (0.125) | (0.222) | (0.513) | (0.536) |
| Marital status of household head (base cat: Unmarried) |  |  |  |  |  |  |  |  |  |  |
| Married | 0.358*** | -0.0880 | -0.176* | 0.106 | 0.847*** | 0.713** | 0.157 | 0.322 | -1.744** | -0.477 |
|  | (0.103) | (0.0887) | (0.0940) | (0.109) | (0.278) | (0.295) | (0.103) | (0.476) | (0.730) | (0.529) |
| Divorced and others | 0.247** | 0.0748 | -0.00656 | 0.116 | 0.789*** | 0.461* | 0.190** | 0.949** | -0.412 | -0.102 |
|  | (0.0966) | (0.0844) | (0.0819) | (0.0850) | (0.259) | (0.262) | (0.0806) | (0.431) | (0.554) | (0.489) |
| Size of household (base cat: 0-2) |  |  |  |  |  |  |  |  |  |  |
| 3 - 5 members | 0.581*** | 0.296*** | 0.393*** | 0.415*** | 0.476*** | 0.901*** | 0.415*** | 1.181*** | 1.184*** | 0.644*** |
|  | (0.0336) | (0.0336) | (0.0383) | (0.0357) | (0.0378) | (0.0968) | (0.0325) | (0.108) | (0.214) | (0.179) |
| 6-10 members | 1.093*** | 0.444*** | 0.714*** | 0.662*** | 0.669*** | 1.124*** | 0.662*** | 1.749*** | 1.588*** | 1.392*** |
|  | (0.0439) | (0.0526) | (0.0547) | (0.0482) | (0.0472) | (0.127) | (0.0469) | (0.146) | (0.294) | (0.248) |
| 10 and above members | 1.656*** | 0.687*** | 0.944*** | 0.795*** | 1.061*** | 1.650*** | 1.206*** | 2.220*** | 1.457** | 1.928*** |
|  | (0.0660) | (0.129) | (0.0972) | (0.0962) | (0.0933) | (0.168) | (0.106) | (0.267) | (0.689) | (0.519) |
| Land holdings in acres (base cat: no land) |  |  |  |  |  |  |  |  |  |  |
| 0.01-0.100 | -0.00154 | 0.109** | 0.120** | 0.0207 | 0.109** | 0.137 | 0.107*** | 0.235** | 0.175 | -0.0481 |
|  | (0.0439) | (0.0517) | (0.0467) | (0.0526) | (0.0505) | (0.129) | (0.0412) | (0.117) | (0.273) | (0.201) |
| 0.101-0..200 | 0.0976* | 0.0968 | 0.133** | 0.102 | 0.176*** | 0.316** | 0.212*** | 0.0204 | -0.219 | -0.141 |
|  | (0.0560) | (0.0742) | (0.0606) | (0.0725) | (0.0641) | (0.158) | (0.0647) | (0.212) | (0.413) | (0.296) |
| 0.201-0.300 | 0.0530 | 0.0539 | 0.150** | 0.0661 | 0.186*** | 0.0400 | 0.113** | 0.454** | 0.364 | -0.0590 |
|  | (0.0644) | (0.0456) | (0.0686) | (0.0545) | (0.0629) | (0.218) | (0.0528) | (0.193) | (0.333) | (0.270) |
| 0.301-.0.500 | 0.114 | 0.0215 | 0.164* | 0.0669 | 0.276*** | 0.134 | 0.154*** | -0.0497 | -0.459 | -0.412 |
| | (0.0757) | (0.0494) | (0.0870) | (0.0898) | (0.0704) | (0.155) | (0.0535) | (0.339) | (0.342) | (0.293) |
| 0.501-1.000 | 0.0588 | 0.107* | 0.173 | 0.0584 | 0.220*** | 0.294 | 0.131** | -0.0609 | -0.485 | -0.518 |
| | (0.0894) | (0.0599) | (0.126) | (0.0913) | (0.0657) | (0.208) | (0.0524) | (0.341) | (0.336) | (0.338) |
| 1.001-3.000 | 0.217*** | 0.185** | 0.426*** | 0.179** | 0.362*** | 0.389** | 0.138** | 0.814* | 0.378 | -0.0536 |
| | (0.0580) | (0.0763) | (0.0641) | (0.0894) | (0.0779) | (0.177) | (0.0590) | (0.461) | (0.516) | (0.435) |
| More than 3.000 | 0.0855 | -0.145 | 0.171 | 0.145 | 0.364*** | 0.168 | 0.143 | -0.0610 | 0.203 | -1.069 |
| | (0.104) | (0.135) | (0.167) | (0.126) | (0.140) | (0.168) | (0.110) | (0.411) | (0.772) | (0.653) |
| Economic activity of household head (base cat: Govt. job) | | | | | | | | | | |
| Private sector job | 0.00885 | 0.0708 | -0.0151 | 0.0246 | 0.0665 | -0.242 | 0.0746 | -0.379** | -0.394 | -0.307 |
| | (0.0492) | (0.0655) | (0.0686) | (0.0544) | (0.0609) | (0.189) | (0.0551) | (0.153) | (0.281) | (0.315) |
| Self employed | -0.0930** | -0.00752 | -0.0236 | -0.0380 | -0.0293 | 0.00946 | 0.0212 | -0.221 | -0.477** | 0.188 |
| | (0.0411) | (0.0509) | (0.0634) | (0.0414) | (0.0626) | (0.183) | (0.0370) | (0.138) | (0.207) | (0.275) |
| Unemployed | -0.0905 | 0.0947 | 0.0452 | 0.158 | 0.206 | 0.291 | -0.121 | -0.577 | -1.919*** | -1.788*** |
| | (0.131) | (0.105) | (0.259) | (0.125) | (0.257) | (0.341) | (0.111) | (0.581) | (0.644) | (0.529) |
| Economically inactive | -0.0160 | 0.00160 | 0.0189 | 0.0172 | 0.0162 | -0.178 | 0.0794 | -0.196 | -0.986*** | -0.658** |
| | (0.0581) | (0.0578) | (0.0720) | (0.0642) | (0.0718) | (0.207) | (0.0537) | (0.169) | (0.284) | (0.324) |
| Constant | 5.658*** | 5.040*** | 5.556*** | 4.463*** | 4.128*** | 4.449*** | 4.001*** | 1.850* | 4.303*** | 2.724*** |
| | (0.228) | (0.153) | (0.169) | (0.196) | (0.815) | (0.796) | (0.244) | (1.027) | (1.426) | (1.024) |
| Observations | 2,762 | 2,762 | 2,762 | 2,762 | 2,762 | 2,762 | 2,762 | 2,762 | 2,762 | 2,762 |
| R-squared | 0.510 | 0.227 | 0.309 | 0.344 | 0.439 | 0.227 | 0.395 | 0.248 | 0.189 | 0.207 |

Robust standard errors in parentheses

*** p<0.01, ** p<0.05, * p<0.1

We also control for the district fixed effect

5656